\documentclass[apj,twocolumn]{emulateapj}

\pdfoutput=1
\usepackage[utf8]{inputenc}
\usepackage{float}
\usepackage{epsfig}
\usepackage{xcolor}
\usepackage{amsmath}
\usepackage{natbib}
\usepackage{graphicx}
\usepackage[breaklinks]{hyperref}

\providecommand{\e}[1]{\ensuremath{\times 10^{#1}}} 

\newcommand{\todo}[1]{
}
\newcommand{\question}[1]{
}
\newcommand{\note}[1]{
}
\newcommand{\checkvalue}[1]{
}

\makeatletter
\newcommand\footnoteref[1]{\protected@xdef\@thefnmark{\ref{#1}}\@footnotemark}
\makeatother

\def \deg {$^{o}$ }

\def \avdal {$A_{V,Dal}$}

\def \avsig {$\sigma_{A_{V}}$}
\def \fred {$f_{red}$}
\def \avfred {\fred{}$\cdot$\avdal{}}
\def \dalcanton {DAL15}

\shorttitle{PHAT XII: Metallicity Distributions in M31}
\shortauthors{Gregersen et al.}

\begin{document}


\title{Panchromatic Hubble Andromeda Treasury XII. Mapping Stellar Metallicity Distributions in M31}
\date{\today}

\author{
    Dylan~Gregersen\altaffilmark{1},         
    Anil~C.~Seth\altaffilmark{1},               
    Benjamin~F.~Williams\altaffilmark{2},    
    Dustin~Lang\altaffilmark{3},             
    Julianne~J.~Dalcanton\altaffilmark{2},   
    Le\'o~Girardi\altaffilmark{4},           
    Evan~D.~Skillman\altaffilmark{5},        
    Eric~Bell\altaffilmark{6},               
    Andrew~E.~Dolphin\altaffilmark{7},       
    Morgan~Fouesneau\altaffilmark{8},        
    Puragra~Guhathakurta\altaffilmark{9},   
    Katherine~M.~Hamren\altaffilmark{9},            
    L.~C.~Johnson\altaffilmark{2},           
    Jason~Kalirai\altaffilmark{10,11},       
    Alexia~R.~Lewis\altaffilmark{2},            
    Antonela~Monachesi\altaffilmark{12},     
    and Knut~Olsen\altaffilmark{13}             
    }


\altaffiltext{1}{Department of Physics and Astronomy, University of Utah, Salt Lake City, UT 84112; {\em \email{dylan.gregersen@utah.edu, aseth@astro.utah.edu}}}

\altaffiltext{2}{Department of Astronomy, Box 351580, University of Washington, Seattle, WA 98195}

\altaffiltext{3}{McWilliams Center for Cosmology, Department of Physics, Carnegie Mellon University, Pittsburgh, PA 15213, USA}

\altaffiltext{4}{Osservatorio Astronomico di Padova - INAF, Vicolo dell'Osservatori 5, I-35122 Padova, Italy}

\altaffiltext{5}{Minnesota Institute for Astrophysics, University of Minnesota, Minneapolis, MN 55455, USA}

\altaffiltext{6}{Department of Astronomy, University of Michigan, 500 Church Street, Ann Arbor, MI 48109, USA}

\altaffiltext{7}{Raytheon, 1151 E. Hermans Road, Tucson, AZ 85706} 

\altaffiltext{8}{MPIA, Koenigstuhl 17, 69117 Heidelberg, Germany}

\altaffiltext{9}{UCO/Lick Observatory, University of California at Santa Cruz, 1156 High Street, Santa Cruz, CA 95064}

\altaffiltext{10}{Department of Physics and Astronomy, Johns Hopkins University, 3400 North Charles Street, Baltimore, MD 21218}

\altaffiltext{11}{Space Telescope Science Institute, Baltimore, MD 21218}

\altaffiltext{12}{MPA, Garching, Germany }

\altaffiltext{13}{NOAO, Tucson, AZ 85719, USA}


\begin{abstract}
    We present a study of spatial variations in the metallicity of old red giant branch stars in the Andromeda galaxy. Photometric metallicity estimates are derived by interpolating isochrones for over seven million stars in the Panchromatic Hubble Andromeda Treasury (PHAT) survey. This is the first systematic study of stellar metallicities over the inner 20~kpc of Andromeda's galactic disk.  We see a clear metallicity gradient of $-0.020\pm0.004$ dex/kpc from $\sim4-20$ kpc assuming a constant RGB age.  This metallicity gradient is derived after correcting for the effects of photometric bias and completeness and dust extinction and is quite insensitive to these effects. The unknown age gradient in M31's disk creates the dominant systematic uncertainty in our derived metallicity gradient. However, spectroscopic analyses of galaxies similar to M31 show that they typically have small age gradients that make this systematic error comparable to the 1$\sigma$ error on our metallicity gradient measurement. In addition to the metallicity gradient, we observe an asymmetric local enhancement in metallicity at radii of 3-6~kpc that appears to be associated with Andromeda's elongated bar.  This same region also appears to have an enhanced stellar density and velocity dispersion.
\end{abstract}



\maketitle



\section{Introduction}
\label{sec:introduction}
    Spiral disks form as a result of complex interactions of star formation, accretion, and gas. These processes are recorded in a galaxy's stellar component. The stellar metallicity, in particular, encodes information on a galaxy's overall chemical enrichment resulting from its history of star formation, gas accretion, and gas outflows.
 
    Spiral galaxies are commonly thought to form ``inside out'' with the inner 
regions forming earliest \citep[e.g.][]{Larson1976}. This effect has been 
observed by measuring a decreasing age with increasing radius in some nearby 
galaxies \citep[e.g.][]{MacArthur2004,Williams2009}.  A metallicity gradient 
is a natural consequence of inside out growth.  Observationally, present-day
metallicity gradients can be measured using HII regions or the atmospheres of young stars.  Studies of the present-day metallicity gradients in the Milky Way and other nearby spirals do show lower metallicities at larger radii \citep[e.g.][]{Rolleston2000,Bresolin2009}. 

    To measure metallicity gradients not at the present-day, but in the
past, one must analyze stellar metallicities as a function of age.  The
simultaneous measurement of age and metallicity is non-trivial, but
can be done using either the observations of resolved stars or using
integrated spectra. Recently, the CALIFA survey \citep{Sanchez2012} used integral field
spectroscopic data to study both the present-day and the mean stellar
metallicity gradients in a large sample of galaxies \citep[][]{Sanchez2014,Gonzalez-Delgado2014,Gonzalez-Delgado2015,Sanchez-Blazquez2014}. The mean mass-weighted stellar metallicity gradient found among the CALIFA galaxies is relatively shallow, $-0.05$~dex/r$_e$, where r$_e$ is the disk effective radius \citep{Sanchez-Blazquez2014}.  The galaxies in this CALIFA subsample are very similar to M31 in Hubble type and luminosity.  Similar results based on photometric modeling were also found by \citep{Roediger2011} for spiral galaxies in the Virgo cluster.  In both studies, there is no clear dependence on Hubble type for spiral galaxies. In the Milky Way, the best gradients utilize measurements of 1000s of individual stars from large surveys (e.g. SDSS) and found a gradient of $\sim$-0.06~dex/kpc \citep{Cheng2012,Hayden2014,Boeche2014,Mikolaitis2014}.

    In addition to the smooth gradients in metallicity, distinct metallicity structures are visible within galaxies. Metallicity substructure in the halos of galaxies have been linked to galactic merger events \citep[e.g.][]{Font2006, Kalirai2006, Ibata2014, Gilbert2014}.  Within the disks of galaxies, structures with distinct metallicity signatures can arise due to bars, bulges and other dynamical structures \citep{DiMatteo2013,Dorman2015}.

    The ability to resolve individual stars all at the same distance makes Andromeda a unique object with which to study massive galaxy disks. Distance is the main source of uncertainty within the Milky Way and large-scale morphology observations are largely restricted to the solar neighborhood. In contrast, stars can be resolved with ease at a wide range of radii in Andromeda.  The Panchromatic Hubble Andromeda Treasury \citep[PHAT;][]{Dalcanton2012} survey recently observed 117 million stars over a third of Andromeda, providing a unique database for studying Andromeda's stellar populations \citep{Williams2014}. 

    In this paper, we present a detailed analysis of the metallicities of Andromeda's red giant branch (RGB) stars using PHAT photometry. Previous work on the RGB metallicities in Andromeda have focused on the halo with limited information in the disk of the galaxy \citep{Bellazzini2003,Kalirai2006,Brown2008,Gilbert2014}.  

    For this study, we assume the following parameters: The center of M31 is at RA=10.68458\deg and Dec=1.2692\deg \citep{McConnachie2005}; M31 position angle and inclination angle are respectively 38\deg and 74\deg \citep{Barmby2006}; the distance modulus is 24.45$\pm$0.05 mag or 776$\pm$18 kpc (mean distance, see \cite{Dalcanton2012} for discussion). Due to the small uncertainty in M31's distance, and the primary dependence of the RGB metallicity on color and not luminosity, the distance uncertainty should have little effect on the analysis we present in this paper. The foreground extinction is $A_V = $ 0.17 mag \citep{Schlafly2011}; M31's effective radius is 8.9$\pm$0.8 kpc \citep{Courteau2011}.

    The paper is structured as follows: In Section~\ref{sec:data}, we explain our data sources, including the RGB star photometry and dust extinction measurements. In Section~\ref{sec:principles-of-metallicities}, we discuss our method for estimating metallicities from isochrone models and the uncertainties caused by photometric crowding and dust extinction. We present and discuss our results on Andromeda's metallicity gradient and a metal rich structure associated with the bar in Section~\ref{sec:results}.

\section{Data}
\subsection{RGB Photometry}
\label{sec:data}
    The PHAT survey observed approximately 117 million stars with the Hubble Space Telescope (HST) in ultraviolet (F275W and F336W), optical (F475W and F814W), and infrared (F110W and F160W) filters using the WFC3/UVIS, ACS/WFC and WFC3/IR HST camera, respectively. The camera pointing angles divided the survey into 23~bricks each consisting of 18~overlapping IR-fields defined from the WFC3/IR camera footprint (for details see \citealt{Dalcanton2012}). 

\begin{figure}[ht]     
    \includegraphics[width=\linewidth]{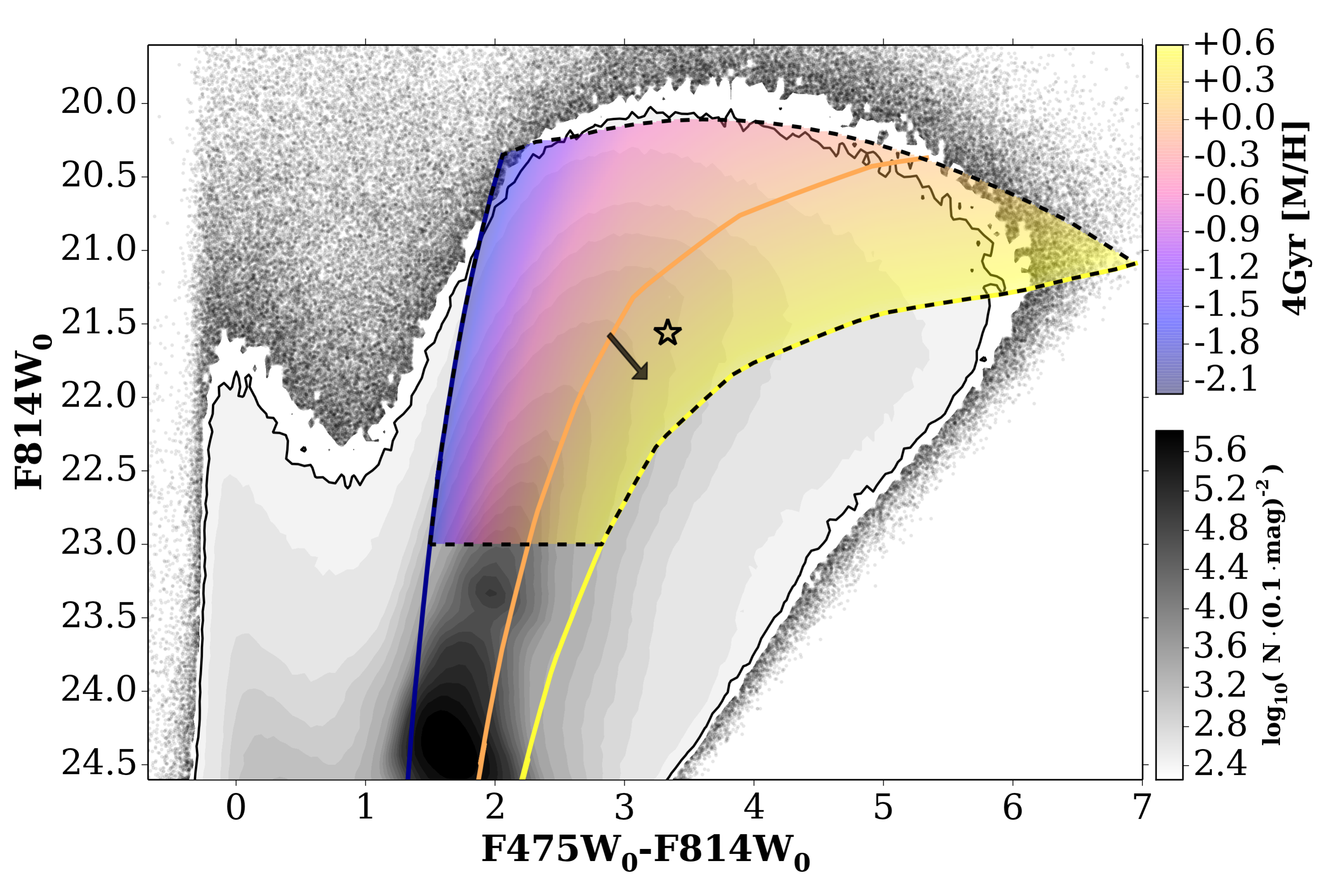}
    \caption{
        Color-magnitude diagram (CMD) of the PHAT optical photometry. Data points and gray scale density contours show our data after photometric quality cuts. The plot bounds contain a total of $\sim$3.4\e{7} stars. The red clump is visible at $F475W_o-F814W_o \approx 1.8$ and $F814W_o \approx$~24.3~mag. The black dashed line indicates the RGB selection box used for this paper (see Section~\ref{sec:completeness-limit}). Color indicates the metallicity. The three color lines plot Padova PARSEC1.2s isochrones at $[M/H]=$-2.0, 0.0, and 0.6~dex (in violet, orange, and yellow, respectively) at a fiducial age of 4~Gyr. The black star marks an example from our data sample at a deprojected radius of 6.1~kpc whose interpolated metallicity estimate is 0.033~dex (see Section~\ref{sec:metallicities}). The black arrow indicates 1~mag of dust extinction in $V$~Band. Extinction causes stars metallicities to be overestimated as discussed at length in Section~\ref{sec:dust-impact}.
    }
    \label{fig:data-selection-cmd} 
\end{figure}

    Two versions of the photometry are publicly available.  First, there is camera-by-camera photometry derived separately on the pairs of filters in each camera producing a UV, optical, and infrared catalog that are then merged together to form a single six-filter catalog \footnote{\label{note:mast-cbyc}Camera-by-camera photometry can be downloaded at http://archive.stsci.edu/prepds/phat/datalist.html} \citep{Dalcanton2012}.  Second, the ``six-filter photometry'' is derived across all filters simultaneously\footnote{\label{note:mast-6filt}Six filter photometry can be downloaded at http://archive.stsci.edu/missions/hlsp/phat/} \citep{Williams2014}. In both versions, each star is parameterized by a position and magnitude, magnitude error, signal-to-noise ratio, crowding, and sharpness for each filter. Sharpness parameterizes the shape of each source relative to the point spread function; and, crowding parameterizes the brightness of nearby sources.
    
    In this paper, we use the optical camera-by-camera photometry to derive RGB metallicity estimates. The primary reason for using the camera-by-camera photometry over the six filter photometry is the availability of extensive artificial star tests (also publicly available from MAST archive). These artificial star tests allow us to understand the photometric completeness and bias (see Section~\ref{sec:artificial-star-tests}). The optical color of RGB stars also is more sensitive to the metallicity relative to the IR photometry, while the UV photometry suffers from significant incompleteness for even the brightest RGB stars.  The smaller effect of dust extinction on the IR data makes it logical to include this data in our analysis, however, artificial stars are only available for the two-camera photometry in the IR, and this data is significantly shallower and crowding limited than the optical data \citep[see Figure~11 of][]{Dalcanton2012}.  The six filter photometry is significantly deeper in the IR due to the ability to use the optical data to determine the positions of stars, but it is not computationally feasible to run artificial star tests over the full PHAT footprint for the six band photometry \citep{Williams2014}.
 
    
    We create an optical only ($F475W$ and $F814W$) photometric catalog with uniform coverage from the individual ACS/WFC field images. This involves two primary steps: defining non-overlapping spatial regions, and filling in the chip gaps of each individual field.  The boundary of each field is first defined using a grid of non-overlapping regions that roughly follow the individual WFC3/IR field footprints.  We additionally crop Bricks 9 and 12, which overlap with neighboring bricks, because of orientation differences described in Section 3.5 of \cite{Dalcanton2012}. The PHAT tiling scheme allows us to use neighboring ACS/WFC fields to fill in the chip gap of each ACS/WFC field. Thus, our final catalog has uniform spatial coverage over the full PHAT survey region but does not make use of multiple overlapping observations.  This same catalog creation process is also used to deal with chip gaps in the the six-filter photometry as described in \cite{Williams2014}.  
    
    
    We apply quality cuts on our photometry. The cuts use signal-to-noise, sharpness, and crowding in the F475W and F814W filters for each star. We use the same parameterization which defines ``gst'' (good star) cuts from \cite{Dalcanton2012} by applying all of the following equations:

$
\begin{array}{rcl}
    F475W\_SNR & \geq & 4.0 \\
    F814W\_SNR & \geq & 4.0 \\
    (F475W\_SHARP + F814W\_SHARP)^2 & \leq &  0.075 \\
    F475W\_CROWD + F814W\_CROWD & \leq & 1.0\\
\end{array}
$ \\              

    \noindent The variables correspond to the names given in the data files. For example, $F475W\_SNR$, $F475W\_SHARP$, and $F475W\_CROWD$ respectively correspond to the signal-to-noise, sharpness, and crowding parameters for the $F475W$ filter.  

    We correct the magnitudes in $F475W$ and $F814W$ for foreground extinction using $A_V = $0.17~mag \citep{Schlafly2011}. This extinction in $A_V$ is converted to $F475W$ and $F814W$ extinctions by multiplying with $A_{\lambda}/A_V$ equal to 1.19119 and 0.60593, respectively\footnote{$A_{\lambda}/A_V$ taken from {\tt http://stev.oapd.inaf.it/cgi-bin/cmd}}. The corrected photometry is indicated by a subscript zero (\emph{e.g.} $F814W_o$). 
          
    In addition to the ``gst'' cuts, we require that $F814W_o \leq 23$ to remove stars with higher photometric bias and lower completeness. The foreground reddening correction is applied before this brightness cut. After applying the spatial and photometric cuts our sample contains $\sim$7.0\e{6} stars.
           
    In Figure~\ref{fig:data-selection-cmd}, we present the optical color-magnitude diagram after quality cuts and foreground extinction corrections. This color-magnitude diagram (CMD) shows the stellar density prior to the magnitude cut in gray scale contours. The red-clump and bump of early-AGB stars are both visible as peaks in the density below our selection region. The dashed area shows the selection region in which we determine RGB metallicities.  In this paper, we estimate metallicities for RGB stars from their positions in the CMD.  Isochrones at a range of metallicities are shown in Figure~\ref{fig:data-selection-cmd}, showing that stars of different metallicities are clearly separated in color; we use this information to infer metallicities after considering a number of subtleties that can affect this measurement (Section~\ref{sec:principles-of-metallicities}). 

\subsection{Sources of Dust Extinction}
\label{sec:dust-sources}
    Dust extinction has a significant impact on the observed photometry and thus our metallicity estimates. We discuss this in more detail in Section~\ref{sec:dust-impact}.  In this section, we describe the source of our dust extinction measurements. 
    
    We use extinction measurements from \cite{Dalcanton2014-dust} (hereafter \dalcanton{}). \dalcanton{}  uses a novel method to directly measure the dust extinction from the IR photometry (filters $F110W$ and $F160W$) of PHAT RGB stars. In the IR CMD, unreddened RGB stars form a tight locus that is insensitive to age and metallicity. Dust extinction causes RGB stars to redden off this locus and form a second, broad RGB.  \dalcanton{} models the unreddened and reddened stars using three parameters: median extinction (\avdal{}), spread of extinctions (\avsig{}), and the fraction of reddened stars (\fred{}).  This latter parameter takes into account the geometry of the dust relative to the RGB stars; 1-\fred{} stars are assumed to be in front of the dust, and thus unaffected by it.  Maps of each parameter are created with 2'' pixels across the PHAT survey region. Note that the central region of the PHAT survey (Brick 1) is not included in the \dalcanton{}  dust maps due to extreme crowding.  
        
\begin{figure}[ht] 
    \includegraphics[width=\linewidth]{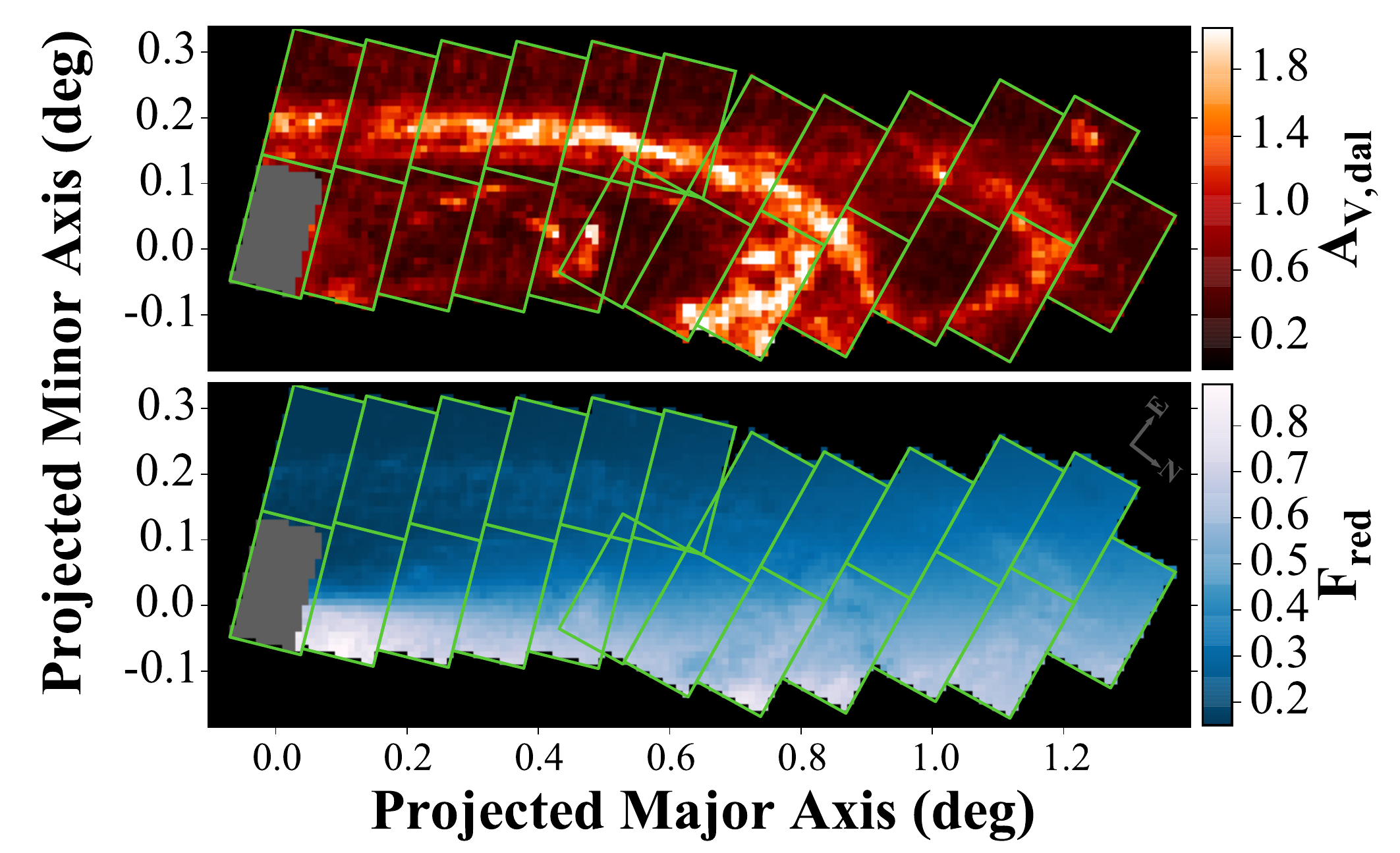}
    \caption{
        Dust maps of M31. \emph{Top} - the mean dust extinction map from \dalcanton{} which we use throughout this paper. The map shows that the dust extinction is highest in the 10~kpc ring region, with lower extinction values in other star-forming regions. \emph{Bottom} - map of the fraction of RGB stars that are reddened, \fred{} from \dalcanton{}. The \fred{} parameter is important in understanding structure caused by dust; e.g., at a fixed $A_V$ increasing \fred{} means more stars will be extincted.  The final \dalcanton{} dust parameter, \avsig{}, is roughly 0.3 everywhere and is not shown here because its effect on our results is negligible.
    }
    \label{fig:dust-maps}      
\end{figure}

    In Figure~\ref{fig:dust-maps}, we present maps of the \dalcanton{} \avdal{} and \fred{}  used in this paper. We rebinned and down-sampled the data into spatial bins used in our final results (0.01\deg squares). The \avsig{} width is approximately 0.3 everywhere and not shown because it has little bearing on interpreting the effect of extinction. In contrast, the \fred{}  plays a significant role. Increases in \fred{} cause the reddening to be more pronounced at the same mean extinction values. For instance, in the case where \fred{}$<0.5$ and high \avdal{}, the median metallicity remains unchanged as stars in front of the thin dust layer are unaffected while those behind are reddened beyond selection bounds. One feature visible in the map of \fred{} is the increase towards the lower (northwest) edge of the survey. In this region a higher fraction of stars appear behind the dust because of the disk inclination.
           

\section{Principles of Photometric Metallicities}
\label{sec:principles-of-metallicities}
    In this section, we explain our method for deriving metallicity estimates and the sources of systematic and random uncertainty in these estimates. This section is organized as follows.

    In Section~\ref{sec:isochrone-models-and-age}, we discuss our choice of isochrone models, which we then use to interpolate the metallicity of individual stars based on their optical color and magnitude in Section~\ref{sec:metallicities}.  We then spatially bin the individual star metallicities to create a map of median metallicity.  Because of the large number of stars, the random sampling uncertainties are negligible.  We address systematic uncertainties in the median metallicity maps due to photometric bias and completeness (\ref{sec:artificial-star-tests}), dust extinction (\ref{sec:dust-impact}), and the uncertainties in the absolute metallicity scale due to isochrone choices and age (\ref{sec:absolute-metallicities}). We correct the median metallicity map for the systematic effects of photometric bias, completeness and dust extinction and present results from this map in Section~\ref{sec:results}.

\subsection{Isochrone Models and Age Assumptions}
\label{sec:isochrone-models-and-age}
    The intrinsic color of an RGB star depends primarily on its age and metallicity. Thus, given the photometry of a star and an assumed age,  we can use isochrones to estimate a star's metallicity from its color and magnitude. When estimating individual star metallicities (Section~\ref{sec:metallicities}), we break the age-metallicity degeneracy using assumptions about the age of M31's disk motivated by literature estimates. In this section, we discuss our choice of isochrone models and the possible impact of age on our metallicity estimates; this impact is further explored throughout the paper. 
    
    
    For isochrone models, we use Padova PARSEC1.2s isochrones \citep{Bressan2012, Chen2014, Tang2014}. At ages $>$1~Gyr, these isochrone models are available at metallicities ($Z$) between 0.0001 and 0.06 \mbox{($-2.18 < [M/H] < 0.6$~dex} using $Z_{\sun}$=0.0152). We use the OBC bolometric corrections described in \cite{Girardi2008} to convert the isochrones to photometric brightnesses in our optical bands. 
    
    We test dependency on the isochrone models by comparing models from PARSEC to BaSTI \citep{Pietrinferni2007} and find the later has a typical offset of approximately +0.3~dex. This offset drops to $\sim$0 for metal poor ($\lesssim-1.0$~dex) and metal rich ($\gtrsim0.3$~dex) stars. Because the majority of our stars have the typical offset, the resulting median metallicity map is only affected by a change in the absolute scale of the metallicity. As we discuss in Section~\ref{sec:absolute-metallicities}, absolute metallicity changes do not affect our results.     
        
    The isochrone models for the bright RGB stars have a well known degeneracy between age and metallicity \citep{Worthey1994}. To break this degeneracy we adopt a single age motivated by the literature. We assume that this age is either uniform throughout the disk, or that it varies linearly with radius.  
    
    Current constraints on the age of M31's disk are relatively minimal.  Spectroscopic measurements of the age by \citet{Saglia2010} show a mean age of $\sim$12~Gyr throughout the bulge, but this drops to $\sim$8~Gyr at their maximum radius of $\sim$2~kpc (500'') where the disk light starts to contribute significantly \citep{Dorman2013}. Only weak constraints on the ancient star formation history (SFH) can be obtained from the PHAT data in the inner disk, but measurements by Williams~et~al.~(~{\em in prep}) suggest a mean SFH age of 10~Gyr throughout the outer disk. We can also get some insight into the age and age gradient present in M31's disk by examining similar galaxies.   As noted in the introduction, the subsample of galaxies in the CALIFA survey analyzed by \citet{Sanchez-Blazquez2014} includes a large number of galaxies similar to M31.  Taking the 25 galaxies within one magnitude of M31 in $r$ band luminosity \citep[CALIFA magnitudes taken from][and use $M_r = -21.5$ for M31]{Walcher2014} and with Hubble type Sab to Sbc, we find that the minimum mass-weighted age in the disks of these galaxies is 3~Gyr, with typical values of $\sim$7~Gyr.  These measurements were made over similar physical scales (0.5-2 $r_{eff}$) as the region we probe with the PHAT data. These old mass-weighted ages of spiral disks are also consistent with previous work by \citet{MacArthur2009}.
    
    While these previous studies suggest that the disk of M31 is old, the average age of stars observed along the upper RGB is not equivalent to the mass-weighted average age of a stellar system due to varying RGB lifetimes and the difference in masses of RGB stars with age.  The average age of stars along the upper RGB is normally {\em younger} than the mass-weighted average stellar age. By simulating a CMD using the star formation history derived by  Williams~et~al.~({\em~in~prep}), we calculate the mean age of RGB stars to be $\sim$4~Gyr throughout the disk.
    
    For the primary median metallicity maps we assume a flat fiducial age of 4~Gyr for our RGB stars. In Section~\ref{sec:fiducial-age-effect}, we show the metallicity gradient is not dependent on the exact fiducial age because the primary effect is a scaling of the absolute metallicity which leaves the gradient unchanged. In Section~\ref{sec:radial-age-effect}, we show that the median metallicity gradient can be changed if we assume a significant age gradient in the disk.

\subsection{Estimating RGB Star Metallicity from Isochrones}
\label{sec:metallicities}

    We estimate individual star metallicities by interpolating isochrones at the distance of Andromeda ($m-M=24.45$) in the $F475W_o-F814W_o$ vs $F814W_o$ color-magnitude diagram. For interpolation, we use the LinearNDInterpolator function within the Python package scipy. This function finds the closest three isochrone points that create a triangle containing a star's color and magnitude using Delaunay triangulation. Then it calculates the star's metallicity by linearly interpolating the metallicity at those three nearest isochrone points. The isochrones used in this procedure are sampled at $\Delta Z = 0.0001$ over the full range of metallicities considered, so the interpolation in metallicity is very small.  In Figure~\ref{fig:data-selection-cmd}, we overplot the star metallicity on top of the stellar density contours in color and magnitude. 

\begin{figure}[ht]     
    \includegraphics[width=\linewidth]{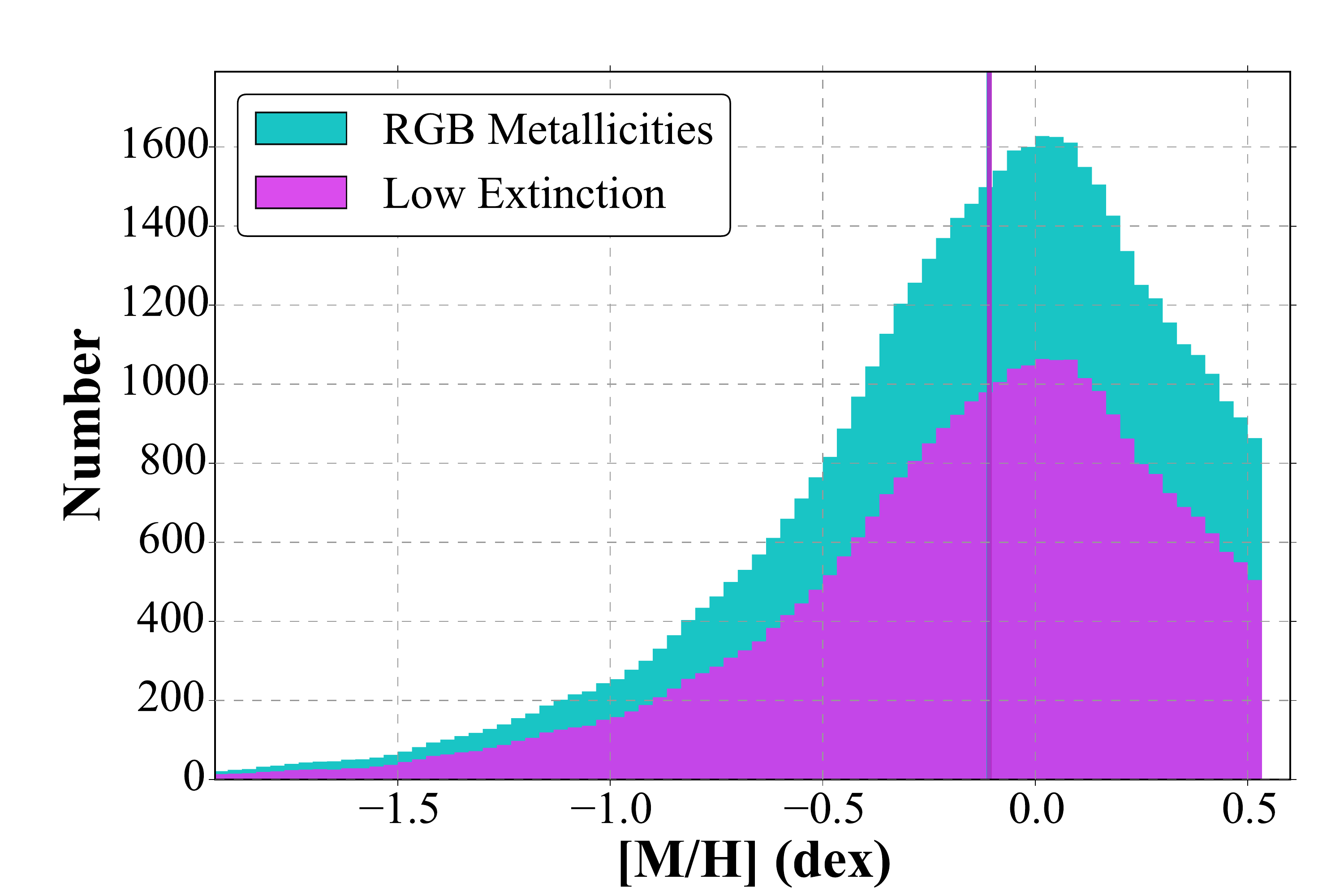}
    \caption{
        The metallicity distribution function (MDF) of our RGB star sample using a flat fiducial age of 4~Gyr. The cyan MDF uses this paper's RGB selection. In purple, we show only RGB stars from low extinction regions within the PHAT survey (\avfred{} $< 0.25$ mag; see Section~\ref{sec:dust-sources}). The median metallicity for both histograms is $\sim$-0.11~dex (shown as a vertical line).   We note this MDF is not corrected for the systematic biases in the metallicity discussed in Section \ref{sec:artificial-star-tests} and \ref{sec:dust-impact}, however, after correction, the MDFs look qualitatively very similar.
    }
    \label{fig:mdf-full}     
\end{figure}

    In Figure~\ref{fig:mdf-full}, we present the metallicity distribution function (MDF) of our RGB stars using an fiducial age of 4~Gyr. We additionally show the MDF generated from stars in low extinction regions. From this figure, we see that the dust has a small effect on the overall MDF of our sample. 

\begin{figure}[ht]     
    \includegraphics[width=\linewidth]{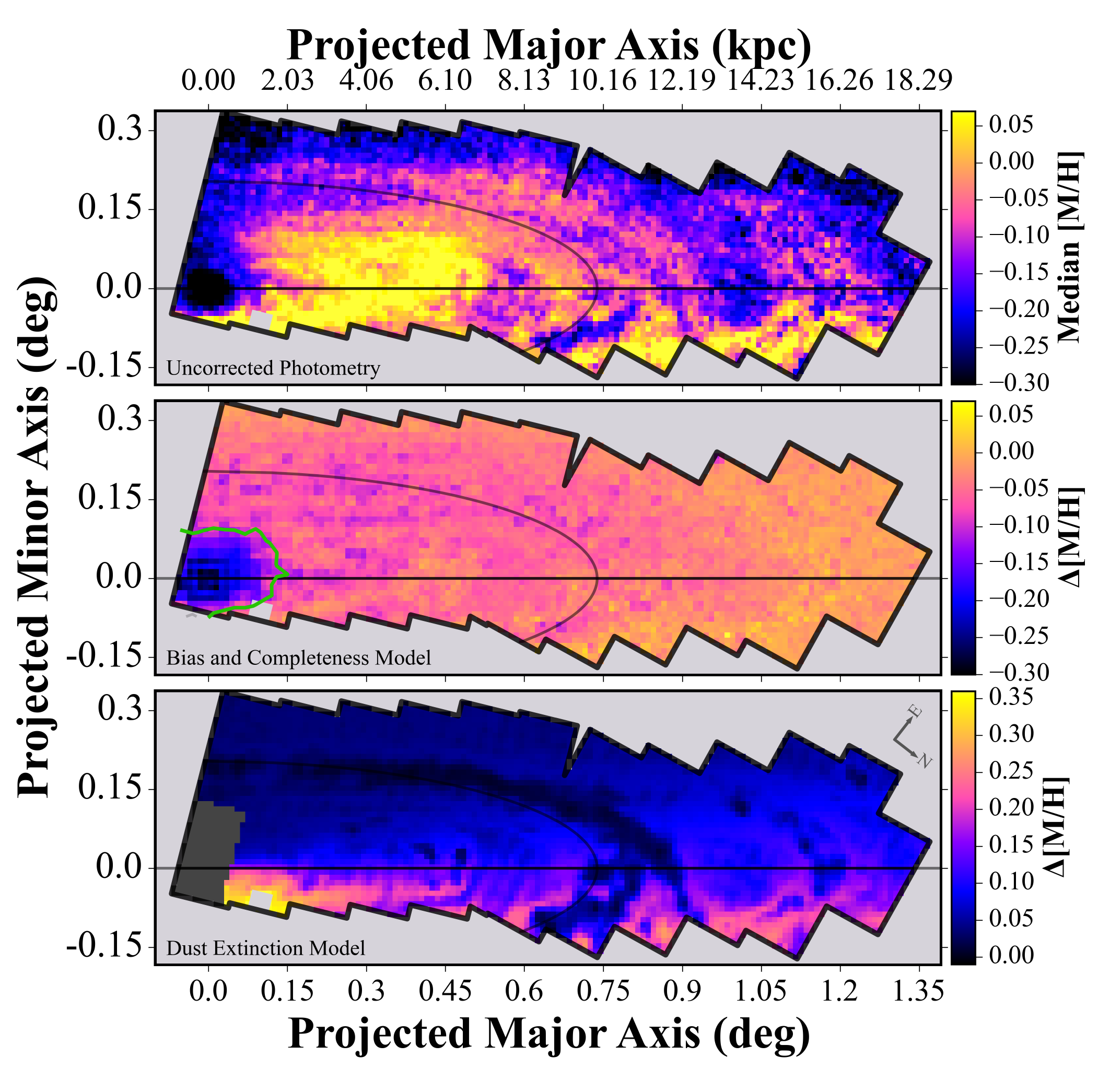}
    \caption{
    {\em Top}: Raw map of the RGB median metallicity without any systematic corrections for dust extinction or photometric bias and completeness. Each spatial bin is 0.01\deg (36~arcsec, 135~pc) square. {\em Middle}: Model for median metallicity changes due to the photometric bias and completeness. These systematics increase towards the central dense regions. The green contour shows the region outside which completeness effects are correctable (see Section~\ref{sec:completeness-limit}). {\em Bottom}: Model for the expected changes in the median metallicity due to dust extinction. We interpret the increases and decreases in median metallicity in Section~\ref{sec:dust-impact}. For low extinction regions the dust causes an increase of $\sim$0.03~dex in the absolute metallicity  and no significant spatial variations in the median metallicity. In our results (Figure~\ref{fig:corrected-median-metallicity-map}), we subtract the two models from the raw median metallicity map to create a corrected version (presented in Figure~\ref{fig:corrected-median-metallicity-map}). Note, all color bars have same range so relative differences can be assessed by eye. 
     }
   \label{fig:uncorrected-med-metal}  
\end{figure}

    Our primary goal in this paper is to examine the spatial variation in the metallicity distribution across Andromeda's disk.  We present a spatial map of the median metallicity of the RGB in the top panel of Figure~\ref{fig:uncorrected-med-metal}. We derive the median metallicity in 0.01\deg (135~pc, projected) square bins. Each bin contains between 100 and 19000 stars. The upper limit becomes 3300 after we exclude the inner regions ($\sim4$~kpc) because of completeness (discussed in Section~\ref{sec:completeness-limit}). The typical bin contains $\sim$560~stars. In addition to the true variations in median RGB metallicity, this map shows features due to photometric bias, completeness, and dust.  We attempt to correct these effects in subsequent sections.

\subsection{Photometric Bias and Completeness Effects on Metallicity}
\label{sec:artificial-star-tests}
    Photometric bias and completeness affect a star's observed photometry and thus the inferred metallicity estimate. Artificial stars allow us to evaluate and ultimately correct for these two systematics. \citet{Dalcanton2012} describes the optical camera artificial star tests in detail. To summarize, $10^5$ artificial stars were distributed uniformly over each PHAT ACS/WCS field of view, resulting in a total of $\sim$4\e{7} stars over the full survey. The input $F475W-F814W$ colors and $F814W$ magnitudes for the artificial stars were split into two halves: the first half sample the full color and magnitude range of the data, while the other half sample the local observed color-magnitude diagram with an extrapolation to the faintest observed magnitudes. Each artificial star was inserted into and recovered from the data images independently using the DOLPHOT package \citep[an updated version of the software presented in][]{Dolphin2002}. The same photometric cuts described in Section~\ref{sec:data} were applied to the output photometry.  We use these artificial stars to measure the photometric bias in our measurement in Section~\ref{sec:bias}, and to model the completeness in Section~\ref{sec:completeness}. Then in Section~\ref{sec:phot-corrections}, we describe how we apply corrections for bias and completeness to our photometric metallicity estimates.


\subsubsection{Evaluating Photometric Bias}
\label{sec:bias}
   Unresolved and bright stars in crowded regions cause a bias in the measured photometry that becomes increasingly large in more crowded regions. This bias generally causes a star to appear brighter and with a color closer to the mean color of the overall galaxy due to blending with fainter undetected stars. 

\begin{figure}[ht] 
    \includegraphics[width=\linewidth]{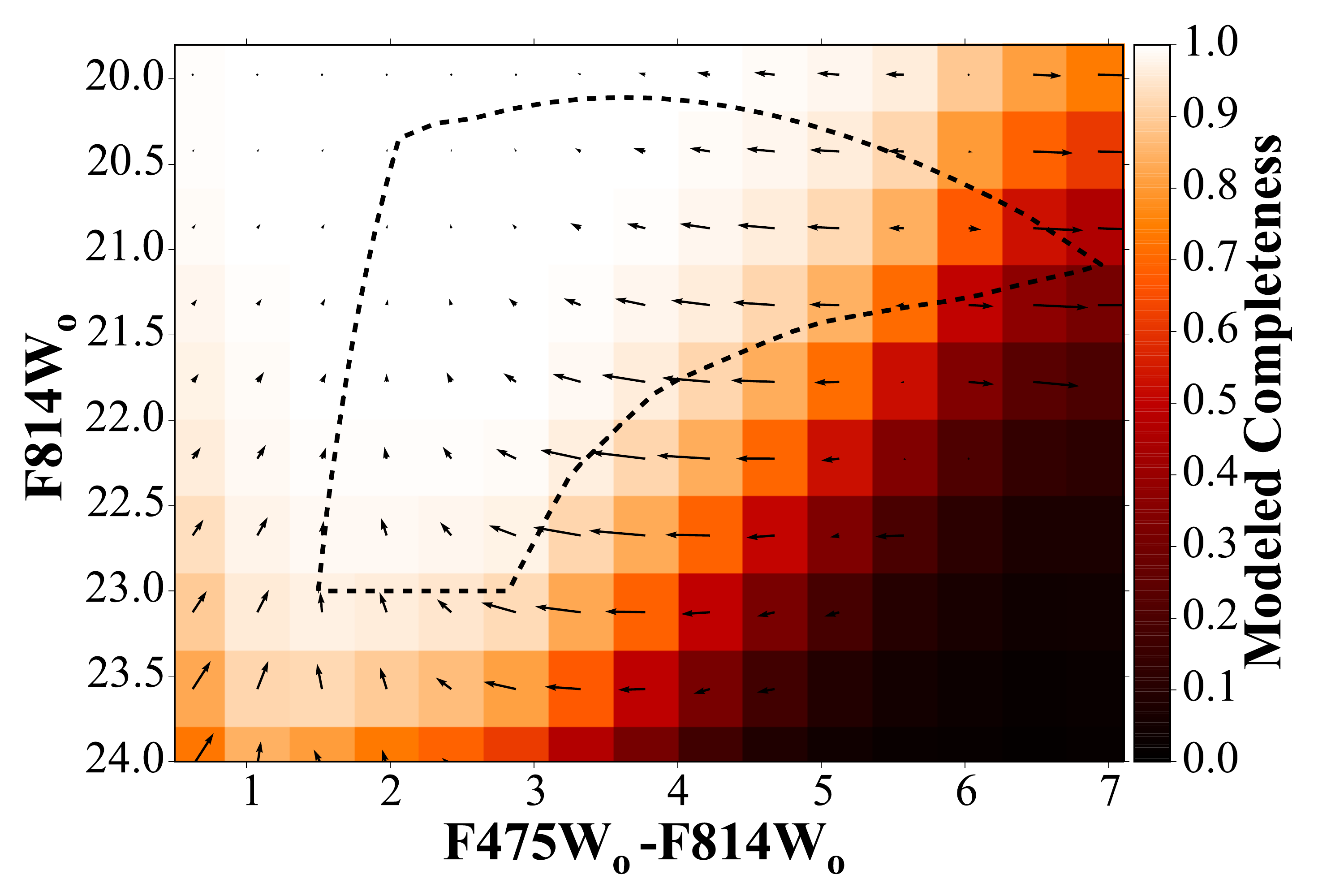}
    \caption{
        Color-magnitude diagram (CMD) showing the mean completeness and bias. This figure is computed from artificial stars at radii of $\sim$2-4~kpc, and therefore highlights the region most affected by completeness and bias that we consider in our analysis (Section~\ref{sec:completeness-limit}). Our RGB selection box is indicated as a dotted line. Each bin is 0.45 mag on each side. The completeness is the number of detected artificial stars divided by the number of those inserted in each CMD bin (Section~\ref{sec:completeness}). The arrows represent the magnitude and direction of bias for each bin (Section~\ref{sec:bias}). The bias generally shifts stars brighter and closer to the median color of the galaxy. We correct our median metallicity determinations for completeness and bias as discussed in Section~\ref{sec:phot-corrections}.
    }
    \label{fig:cmd-bias-completeness} 
\end{figure}

    In Figure~\ref{fig:cmd-bias-completeness}, we show the average color and magnitude bias in bins across the CMD for a region of M31 at radii between 4-5~kpc where the bias is significant. The bias of each individual artificial star is measured from the recovered magnitude minus the input magnitude. Because the star's recovered CMD position is changed from its original position, this photometric bias results in a bias in our metallicity estimates as well. Examining Figure~\ref{fig:cmd-bias-completeness} shows stars within our RGB region are typically detected to be brighter and bluer than their intrinsic CMD position.  Our metallicity estimates for these stars are thus biased lower than the true metallicity, with the effect being more severe for high metallicity stars which are closer to the $F475W$ detection limit due to their intrinsically redder colors.  
   
    In Section~\ref{sec:phot-corrections}, we use the measured bias to partially correct for this effect. For computational efficiency, we create a binned data set of the average bias. We bin using 4 dimensions: two spatial dimensions with step size of 0.02\deg (270~pc) in both projected major and minor axis of M31; and, two color-magnitude dimensions with step size of 0.45 mag in both dimensions (sizes for spatial and CMD bins seen in Figures~\ref{fig:min-completeness} and \ref{fig:cmd-bias-completeness} respectively). The result is a median of 3.3\e{4} stars input for every spatial pixel and 36 input stars for every 4D bin.  In each bin with $>10$ recovered artificial stars, we calculate the mean photometric bias for each filter from all the artificial stars. Stars in 4D~bins with $<10$ artificial stars receive no correction value. 

              
\subsubsection{Evaluating our Photometric Completeness}
\label{sec:completeness}
    The completeness is the number of detected stars divided by the total number of stars present at a given magnitude and color. Our completeness correlates with magnitude; at fainter magnitudes we detect a smaller fraction of stars and are thus less complete.  Because we require stars to be detected in both filters, each filter contributes to the completeness; for blue stars the F814W magnitude determines the completeness (thus creating a horizontal completeness limit on our CMD), while for redder stars, the F475W magnitude limits the completeness, creating a diagonal completeness limit on our CMD. These completeness limits are visible in Figure~\ref{fig:cmd-bias-completeness} where the darkest pixels represent $\sim0\%$ completeness. The faint F475W magnitudes of metal-rich giants causes these stars to be most affected by completeness, and thus we detect a smaller fraction of the most metal-rich stars.  

    We measure the completeness from the artificial star tests as follows. We define an artificial star as detected if it passes the photometry cuts defined in Section~\ref{sec:data} and has an absolute magnitude change less than 0.75~mag. The completeness is then simply the fraction of artificial stars detected in a particular spatial and CMD volume. We measure the completeness using the same 4D bins described previously in Section~\ref{sec:bias}.
   
    To correct for completeness, we divide the total number of artificial stars recovered and inserted at a given CMD position using the {\em input} magnitudes (i.e. after the bias correction, Section~\ref{sec:phot-corrections}). This is somewhat different from the standard practice where completeness is calculated using the observed magnitudes to determine the detections and the input magnitudes to determine the total number of stars inserted at a given CMD position.  
   
    Due to the small number of stars in each individual 4D bin, the errors on the completeness in an individual bin are large.  To reduce these uncertainties, we model the completeness at a given color as a function of $F814W_o$ magnitude using a sigmoid function (e.g. \citealt{Bik2013}):
        
\begin{equation} 
    \label{eq:completeness-model}    
    C_{model}(F814W_o|a,b) = (1-exp(\frac{1}{a}\cdot{}F814W_o+b))^{-1}
\end{equation}

\noindent Our modeling process is a loop with five steps:
    First, we select artificial stars within a 2-dimensional spatial bin. Second, we select artificial stars within a particular color bin and in the neighboring color bin(s). Using neighboring bins increases the number of stars used for each fit. Also as a result the fit parameters become covariant. This is beneficial because the covariance effectively smooths along the color dimension and makes the measurement more robust. Third, we bin the color-selected stars into independent $F814W_o$ magnitude bins. The completeness for each 2-dimensional color-magnitude bin is the fraction of detected to input artificial stars.  An error in this measurement is also estimated. Fourth, we fit the completeness vs $F814W_o$ to Equation~\ref{eq:completeness-model} using $\chi^{2}$ minimization and determine $a$ and $b$ for that color bin. Fifth, we use the model to assign a fitted completeness to each 4D bin. 
    
    At the end of this process, we have robust measurements of our completeness binned in 4D.  We show a spatial map of our completeness for RGB stars in Figure~\ref{fig:min-completeness}; this figure is discussed further in Section~\ref{sec:completeness-limit}.


\subsubsection{Applying Photometric Corrections}
\label{sec:phot-corrections}

    Using the results from \ref{sec:bias} and \ref{sec:completeness} We correct our final median metallicity measurements for the photometric bias and completeness systematics. To apply this correction, we first correct the observed photometry for photometric bias. Then, when aggregating into the median metallicity, we correct for completeness by weighting each star by the inverse of the recovered completeness from Section~\ref{sec:completeness}. In this section, we further describe these corrections and the creation of a photometrically corrected median metallicity map. 
       
    To correct the observed star photometry for the photometric bias in each filter, we first bin the observed stars using the same 4D bins described in Section~\ref{sec:bias}. Then, we select each spatial bin and consider only the mean bias values in the associated CMD bins.  We interpolate the bias for each star in both filters using the CMD bin centers and mean bias values (as we did for the metallicity in Section~\ref{sec:metallicities}). Finally, the observed star photometry is corrected for bias by subtracting interpolated bias from the observed photometry.  
    
    We use the bias corrected photometry to estimate a photometric completeness for each star. The process is similar to interpolating the bias values for each star, only we now use the 4D recovered completeness data from Section~\ref{sec:completeness}. We apply a minimum completeness limit of 0.05 so when weighting by completeness no star can dominate the calculations. This limit affects $<0.0002\%$ of our stars.
    
    For every spatial bin in our final maps, we calculate a bias and completeness corrected median metallicity. We first re-derive metallicities for the stars using the bias corrected photometry. To calculate the median metallicity, we weigh each star by the inverse of its completeness and then take the 50th percentile of the stars' MDF. 
   
    The middle panel of Figure~\ref{fig:uncorrected-med-metal} shows the effect of bias and completeness on the median metallicity map; specifically, it shows the median metallicity determined from the uncorrected photometry minus the median metallicity after we apply our bias and completeness correction.  The bias and completeness correction is largest near the center where the density of stars is high and is negligible in the outer parts of the galaxy. We discuss the limits to these corrections in Section~\ref{sec:completeness-limit}. 
     
    This correction model is not exactly accurate due to uncertainties in the artificial star tests caused by stochasticity and assumptions in the prior distribution of the stars. Nevertheless, the model provides a reasonable and robust correction to this known systematic.

\subsubsection{Completeness Limit}
\label{sec:completeness-limit}
    Where the stellar density becomes high, photometric bias and completeness effects become large. When the completeness drops to near 0\%, our bias and completeness corrections are no longer valid as they require the intrinsic distribution of stars to be at least partially detected. This limits how close to Andromeda's galactic center we can measure metallicities and limits our ability to detect the most metal-rich stars.
      
\begin{figure}[ht]     
    \includegraphics[width=\linewidth]{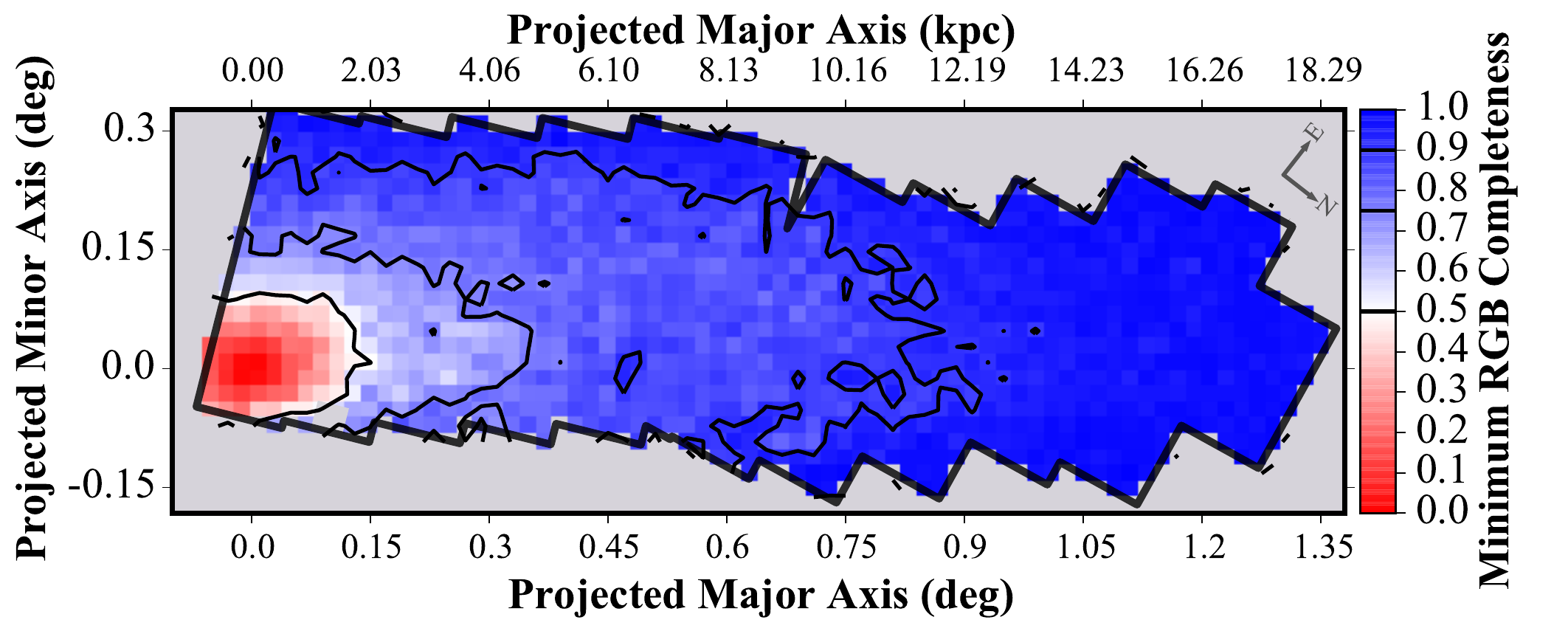}
    \caption{
        Map of the minimum completeness for stars within our RGB selection box (see Figure~\ref{fig:data-selection-cmd}). This metric is calculated by assigning a completeness to all the stars then taking the minimum from the 90th percentile---in CMD density---of our data.  The solid lines indicate 50\%, 75\%, and 90\% minimum completeness. In general, the stars most affected by completeness are the reddest, highest metallicity portion of the RGB.  At the center of Andromeda these stars are completely undetectable. As the sample becomes more incomplete it becomes impossible to correct properly for the bias and completeness; we therefore only analyze the area with $>50$\% completeness (see Section~\ref{sec:completeness-limit}).
    }   
    \label{fig:min-completeness} 
\end{figure}

    We use a minimum completeness metric to quantify exactly how close to the center our corrections are applicable. At each spatial position, we measure the completeness along the 90th percentile density contour of our RGB stars (shown as a black contour line on Figure~\ref{fig:data-selection-cmd}) limited our RGB $F814W$ magnitude limit. We use the minimum completeness along the contour to create the map shown in Figure~\ref{fig:min-completeness}. For almost all spatial positions the minimum completeness is measured at the most red RGB tip. This means the completeness along our $F814W$ magnitude limit is greater than this minimum completeness metric at practically all spatial positions.

    From the minimum completeness map (Figure~\ref{fig:min-completeness}), we define the limiting central region at the 50\% contour which is $\sim$4~kpc.  At this limit, the bias and completeness correction is $\sim$0.1~dex. We exclude data within from our final analysis. Beyond the limit, we find the same correction has a $\sim1\sigma$ effect on our metallicity graident (Section~\ref{sec:metallicity-gradient}) and no bearing on our observed metal rich bar structure (Section~\ref{sec:metal-rich-bar}). We conclude this limiting region to be a reasonable choice. In all subsequent figures, we include the results at all radii but shade the excluded central region.

\subsection{Effects of Dust Extinction}
\label{sec:dust-impact}
    
    Dust extinction causes stars to appear dimmer and redder. For RGB stars, this causes us to overestimate the metallicity of reddened stars (see arrow in Figure~\ref{fig:data-selection-cmd}). We use information on the dust distribution (see Section~\ref{sec:dust-sources}) to both identify high extinction regions and to model the effect of dust extinction on the median metallicity. 
    
    To identify regions where the extinction is likely to cause biased metallicity estimates, we multiply the mean extinction by the fraction of reddened stars (\fred{}) from \dalcanton{} within each spatial bin. We find \avfred{}~$ > 0.25$~mag is a good limit for identifying regions where extinction significantly biases our results.
    
    We measure the significance of dust extinction on our median metallicity measurements by forward modeling the effect of dust extinction on an unreddened fiducial CMD. 

    To forward model our data, we first need a realistic input CMD. We take this input CMD from unreddened regions in the survey.  For the unreddened fiducial CMD, we use stars in the low extinction region (\avfred{} $<0.25$~mag) at $\sim$~14~kpc along the major axis (visible in Figure~\ref{fig:dust-maps}).  Our final result should be insensitive to low level extinction present in our fiducial CMD because we measure the relative change in median metallicity of the sample. Our results are very similar using other low extinction regions within the survey (e.g. along the NE edge).  The primary weakness of our approach is that we do not account for the stellar population differences (i.e.~young stars) we expect between the dust-free and dusty regions of the galaxy; trying to account for this change would require full star formation history modeling and is beyond the scope of this paper.  We expect that any young contaminants not included in our fiducial CMD would be brighter bluer stars reddened into our CMD region, and thus would likely reduce the median metallicity at moderate extinctions.  
        
    We apply reddening to this CMD to measure the resulting change in median metallicity. The extinction for each individual star in the CMD is drawn from the extinction probability function from \dalcanton{} using their maps of the best fit \fred{}, \avdal{} and \avsig{} dust parameters (see Figure~\ref{fig:dust-maps}). The stars are binned into 0.01\deg spatial bins which correspond to the bin size used in our median metallicity maps. In each bin, we take the associated dust parameters and determine a random extinction for each star. Then we apply that extinction to create a new reddened data set and re-evaluate the metallicity as described in Section~\ref{sec:metallicities}.  Finally, we measure the change in median metallicity between new reddened data and the original unreddened sample.     
    
\begin{figure}[ht] 
    \includegraphics[width=\linewidth]{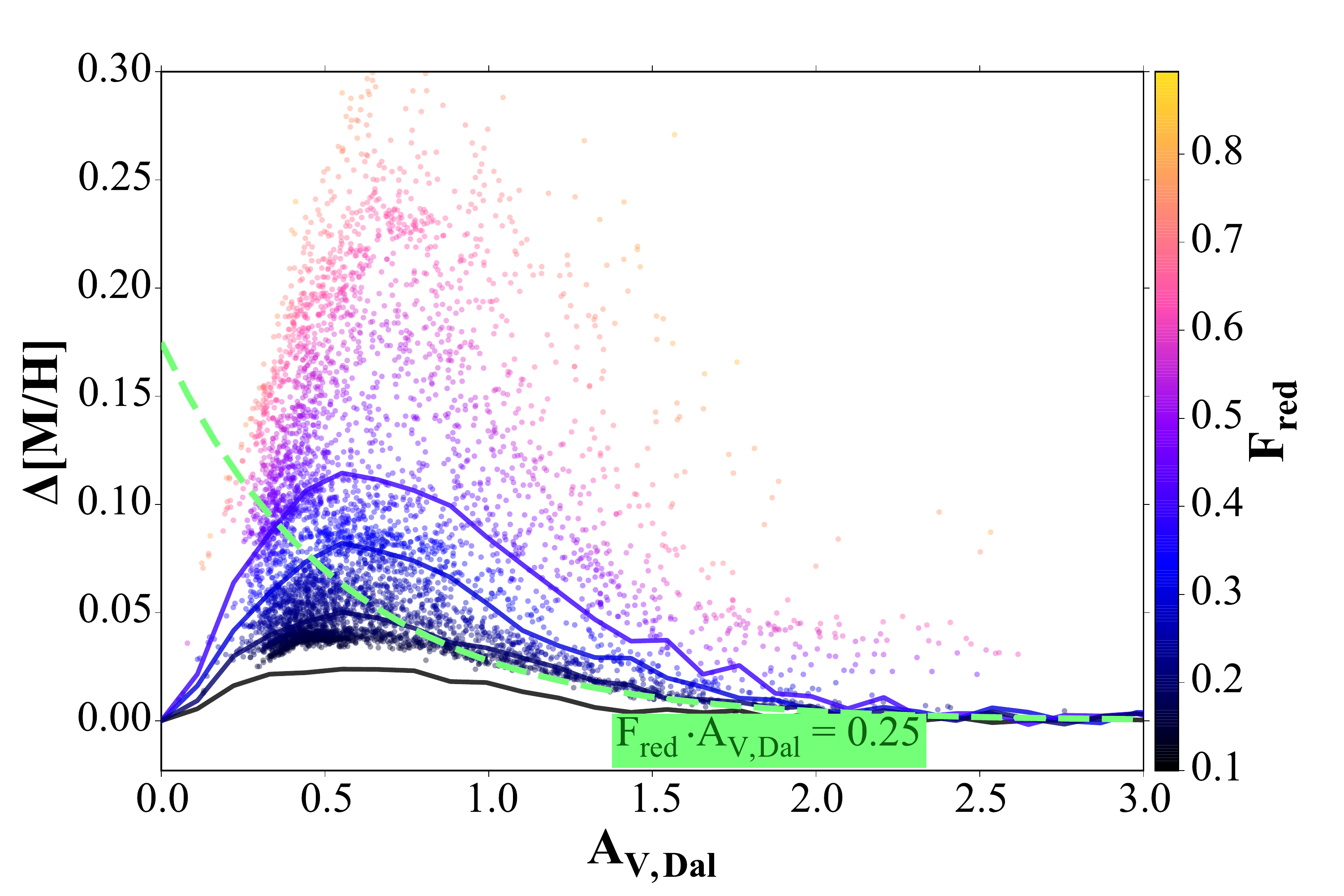}
    \caption{
        Simulated change in median metallicity ($\Delta$[M/H]) for each spatial bin in the data.  The simulation uses a fiducial CMD taken from a low-dust region and applies reddening to individual stars using the dust parameters (\fred{},\avdal{}, and \avsig{}) from the \dalcanton{} dust maps for every spatial bin. The solid lines show changes in median metallicity at constant \fred{} values of 0.1, 0.2, 0.3, and 0.4. Along one of these solid lines (and assuming \avsig{}=0.3). The dust increases the median metallicity until \avdal{}$\approx$0.5 because the reddened stars receive an overestimated metallicity. The median metallicity decreases for high extinction when the reddened stars are moved beyond the edge of our RGB box and no longer affect the measurement. The points below the dashed line are low extinction regions with \avfred{} $< 0.25$.  These are the regions used in our metallicity gradient measurement.  Their $\Delta$[M/H] is $<0.12$~dex, with typical values of 0.03~dex.  Details on the dust simulations are given in Section~\ref{sec:dust-impact}; a map of $\Delta$[M/H] is shown in the bottom panel of Figure~\ref{fig:uncorrected-med-metal}.
    }
    \label{fig:dmedian-vs-dust}          
\end{figure}

    In Figure~\ref{fig:dmedian-vs-dust}, we present the change in median metallicity as a function of \avdal{} for each spatial bin colored by \fred{}.  These two parameters dominate the predicted change in median metallicity estimates; changing \avsig{} has a $<0.005$~dex change in the median metallicity estimates.  As previously noted, for a particular \avdal{} the change in median metallicity increases with increasing \fred{}. Lines of constant \fred{} (and constant \avsig{}=0.3) are also plotted on Figure~\ref{fig:dmedian-vs-dust}. For these constant \fred{} lines, the change in median metallicity first increases then decreases with increasing \avdal{}. The increases are easily understood as stars becoming reddened and appearing more metal rich. The decrease is caused by stars becoming so reddened they are moved beyond our RGB selection box and the most metal rich isochrone; the remaining unreddened stars (which are in front of the dust distribution) yield a very similar median metallicity to the input CMD.
    
    The individual points in Figure~\ref{fig:dmedian-vs-dust} correspond to individual spatial bins. We also present these data as a map of the change in median metallicity (lower panel of Figure~\ref{fig:uncorrected-med-metal}). This map shows sharp changes due to dust especially along the upper (northeastern) edge of the survey and in the 10~kpc ring. Our final results avoid these regions by imposing a cut to remove these high extinction regions (\avfred{}$ > 0.25$~mag). 
    
    In the low extinction regions, the dust still has a low level effect. We apply a dust correction by subtracting the change in median metallicity due to dust. The changes in metallicity are small (typically $\sim0.03$~dex) and---more importantly---distributed randomly over the survey area so the result is only a systematic on the absolute metallicity scale and not on spatial variations. Systematics on the absolute metallicity scale do not heavily impact our results.

\subsection{Uncertainties in Absolute Metallicities}
\label{sec:absolute-metallicities}

    The absolute metallicities we determine have large uncertainties which we discuss in this section.

\noindent {\em Age: } Older populations have redder RGBs than younger populations of the same metallicity.  Therefore, using older fiducial ages results in lower metallicity estimates.  This effect is quite significant, with metallicity estimates varying by $\sim$0.4 dex when changing the age from 3 to 13~Gyr (we discuss this further in Section~\ref{sec:fiducial-age-effect}). Overall, this is the dominant uncertainty in the absolute metallicity.  Because we are primarily interested in the metallicity gradient, an age gradient can significantly affect our result; this is discussed further in Section~\ref{sec:radial-age-effect}.

\noindent {\em Isochrones:} As noted in Section~\ref{sec:isochrone-models-and-age}, use of the BaSTI isochrones results in a median metallicity that is $\sim$0.3 dex higher than the PARSEC~1.2s isochrones we adopt.

\noindent {\em Dust:} From our simulations in the previous section, we find the low level dust extinction present throughout the survey results in a slight overestimate of the intrinsic metallicity; however this bias is only $\sim$0.03~dex.

\noindent {\em Photometric Bias and Completeness:} Near the center, the bias and completeness correction raise our median metallicity by more than 0.1~dex.  Unlike the effects above, this is primarily a bias in the relative metallicity between the center and outskirts, as the bias and completeness have no affect on the median metallicity in the outer disk regions.

    Because of the large uncertainties in the absolute metallicities, we choose to focus on relative measurements for our results. Particularly, the metallicity gradient of Andromeda and metallicity substructure of the disk. After applying corrections for known systematics, we present the median metallicity map and results in the next section.   

\section{Results: Metallicity Gradient and Structures in Andromeda}
\label{sec:results}

\begin{figure*}[tp] 
    \centering 
     \makebox[\textwidth]{
    \includegraphics[width=0.9\paperwidth]{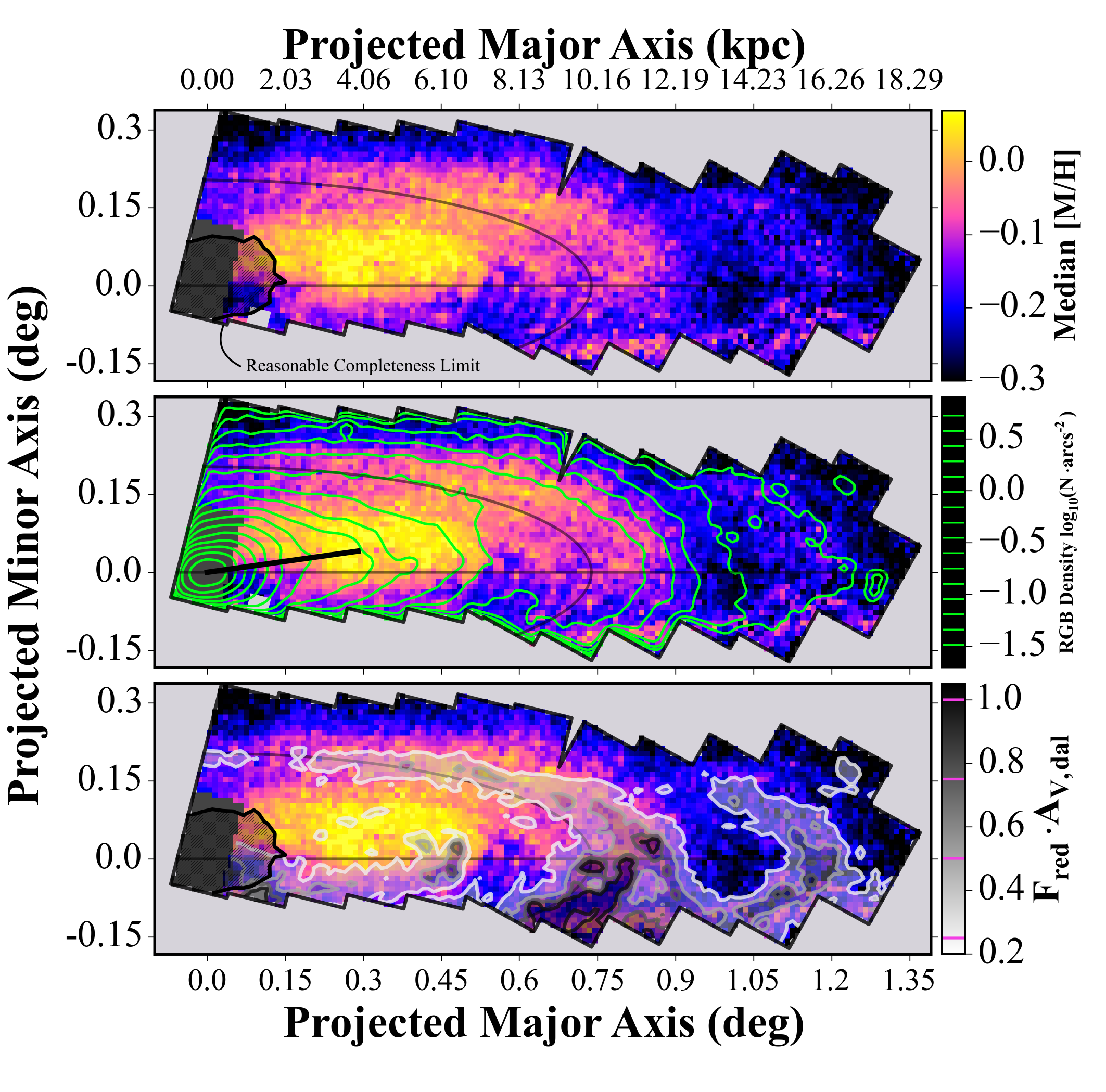}
    }
    \caption[width=0.9\paperwidth]{ 
        {\em Top Panel}~-- Map of the RGB median metallicity after corrections for dust extinction, photometric bias, and completeness. The shaded black regions indicate the crowded central region where our bias and completeness corrections fail in recovering the true metallicity distribution (Section~\ref{sec:completeness-limit}). Included as a reference point, the black ellipse is a ring with with constant deprojected radius at 10~kpc. Our measurement of the metallicity gradient uses these corrected median metallicities (Section~\ref{sec:metallicity-gradient}). {\em Middle Panel}~-- on top of the median metallicity map, we plot green contours at constant RGB density as inferred from the IR photometry (\cite{Dalcanton2012}, Seth~et~al.~{\em~in~prep}). Additionally, a solid black line shows Andromeda's bar length and azimuthal offset measured by \cite{Athanassoula2006}. We observed a region of high median metallicity correlated with the RGB density and bar. In Section~\ref{sec:metal-rich-bar}, we argue these high metallicities are likely a result of bar interaction. {\em Bottom Panel}~-- we overplotted the median metallicity map with the average reddening (\avfred{}). Every spatial bin below the lowest dust contour we define as low extinction (\avfred{}$ < 0.25$~mag). We base our results on these low extinction areas. This figure shows the high metallicity region we associate with Andromeda's bar is not caused by dust.   
    }
    \label{fig:corrected-median-metallicity-map}      
\end{figure*}

    In this section, we analyze the radial and spatial variation of the median RGB metallicity. Our results are based on the spatial map of corrected median metallicities presented in Figure~\ref{fig:corrected-median-metallicity-map}. Our results focus on two interesting features in the corrected median metallicity map. The first is an overall gradient towards lower metallicities at larger radii (Section~\ref{sec:metallicity-gradient}). The second feature is a region of high metallicity at radii of 3-7~kpc lying just above (southeast of) the major axis, which we associate with an elongated bar (Section~\ref{sec:metal-rich-bar}). 
    
\clearpage

\begin{figure}[ht] 
    \includegraphics[width=\linewidth]{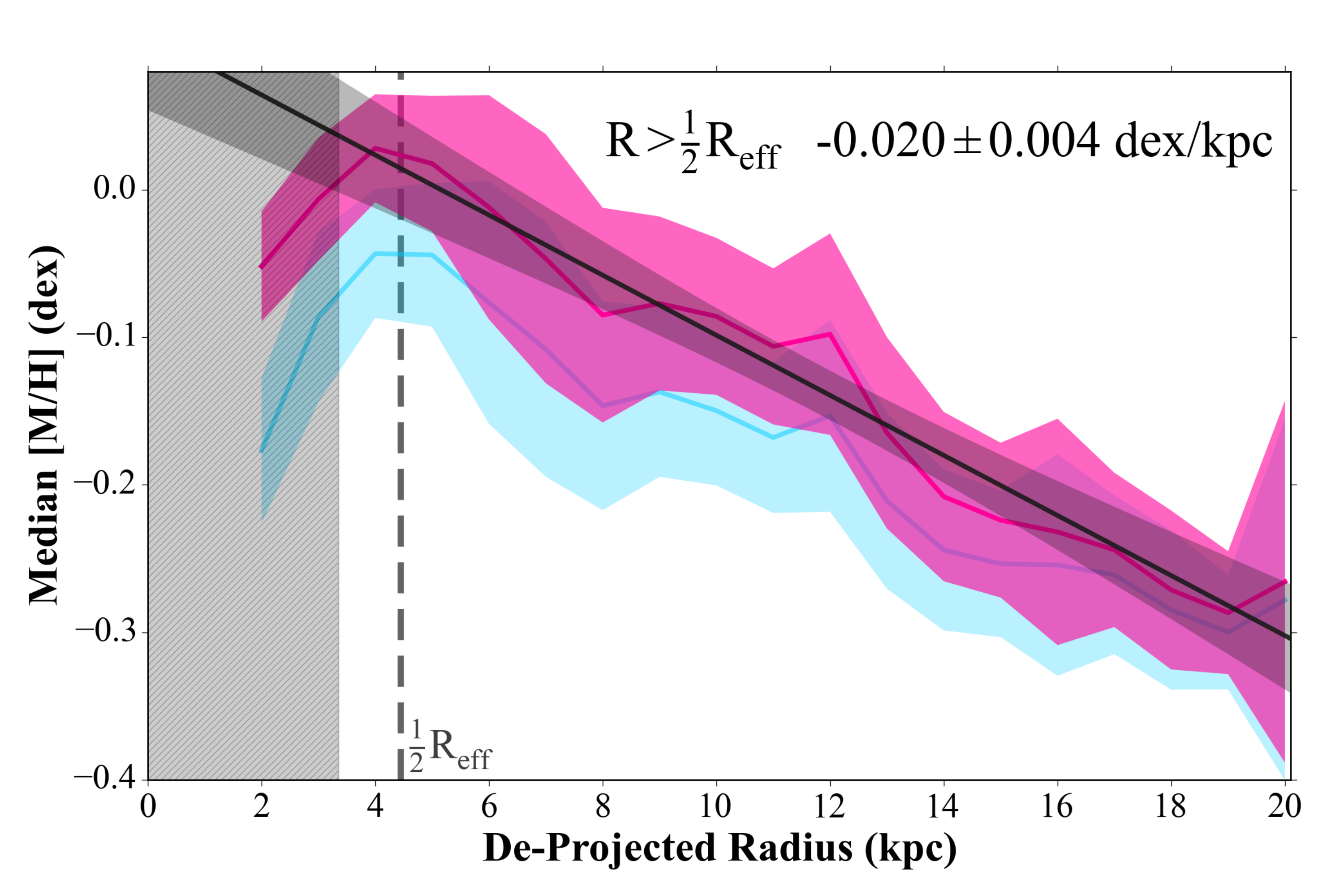}
    \caption{
        The median metallicity gradient in M31.  To create this figure, the median metallicity map (Figure~\ref{fig:corrected-median-metallicity-map}) was divided into annular bins with $b/a=0.275$; within these bins the mean of the map is shown as the solid line, while the shaded region shows the standard deviation of the pixels in the bin.  High extinction regions (\avfred{} $> 0.25$) were excluded. The red line shows the gradient including corrections for dust extinction and photometric bias and completeness, while in cyan the metallicities have only been corrected for dust extinction. The bias and completeness correction changes the gradient by $\sim$1$\sigma$, while the dust extinction correction changes the gradient by less than a 0.001$\sigma$. The black line and shaded region shows the best linear fit at deprojected radii $R > 4.45$~kpc~($\frac{1}{2} R_{eff}$; \cite{Courteau2011}) to the corrected metallicities, the best-fit slope is -0.020$\pm$0.004 dex/kpc.  The black shaded region at $R \lesssim 4$~kpc where our RGB data becomes incomplete at high metallicities.
    }
    \label{fig:feh-vs-radius} 
\end{figure}  


\subsection{Radial Metallicity Gradient}
\label{sec:metallicity-gradient} 
 
    We present the first measurement of the RGB metallicity gradient in the M31 disk from 4 to 20~kpc. We binned the median metallicity (shown in Figure~\ref{fig:corrected-median-metallicity-map}) into elliptical apertures with a size of 1~kpc along the major axis, and assuming $b/a = 0.275$ \citep{Barmby2006} to account for the disk's inclination.  We exclude the high extinction regions with \avfred{}$> 0.25$ to limit the effect of dust on our measurement; this limit is shown as the white contour in the bottom panel of Figure~\ref{fig:corrected-median-metallicity-map}.  After excluding these regions, we take the average and standard deviation of the median metallicity within each elliptical aperture to create the radial metallicity profile shown in Figure~\ref{fig:feh-vs-radius}.

    We fit a gradient to the measured metallicity profile at $R > 4.45$~kpc~($\frac{1}{2} R_{eff}$; \cite{Courteau2011}). This radial cut excludes the regions we distrust due to low completeness (shown as the shaded region in Figures~\ref{fig:corrected-median-metallicity-map} and \ref{fig:feh-vs-radius}). We fit the mean and standard deviation of the metallicity in each radial bin using a Monte Carlo simulation with a linear fit. For each iteration, we calculate the best fit slope and intercept using least squares optimization. We repeat this and obtain a distribution of fitted slopes and intercepts; this procedure also allows us to examine the covariance of the slope and intercept. We show the best fit parameters and 1$\sigma$ contours in Figure~\ref{fig:fiducial-age-params} for various fiducial ages including our primary result with 4~Gyr.

    We measure Andromeda's metallicity gradient to be $-0.020\pm0.004$~dex/kpc. This gradient uses a 4~Gyr fiducial age (see Section~\ref{sec:isochrone-models-and-age}, Fig~\ref{fig:fiducial-age-params}) and our corrected metallicities. The error on the gradient is calculated using the Monte Carlo technique described above; bootstrapping errors are significantly smaller than this quoted value.   The photometric bias, completeness and dust corrections have little impact on producing the observed gradient: the dust corrections change the gradient by $<1\%$, and the photometric bias and completeness corrections steepen the relation somewhat, but still agree at about the 1$\sigma$ level. We also measure a consistent metallicity gradient (within 0.25$\sigma$) when selecting stars just along the major axis which indicates the assumed inclination angle has little impact on the gradient. The intercept (central metallicity) of this fit is [M/H]$=0.11\pm0.05$; though caution should be taken in interpreting the absolute metallicity as explained in Section~\ref{sec:absolute-metallicities}.

    To facilitate comparisons with the literature, we also present our gradient with different units. In terms of Andromeda's half light radius (8.9$\pm$0.8~kpc; \cite{Courteau2011}), the gradient is $-0.181\pm0.040$~dex/$R_{eff}$. Using the radius of the isophote at a bolometric B-band of 25~mag ($21.5$~kpc; \cite{RCS1991}), the gradient is $-0.43\pm0.09$~dex/R25.

\subsubsection{The Effect of Changes to the Fiducial Age}
\label{sec:fiducial-age-effect}

\begin{figure}[ht] 
    \includegraphics[width=\linewidth]{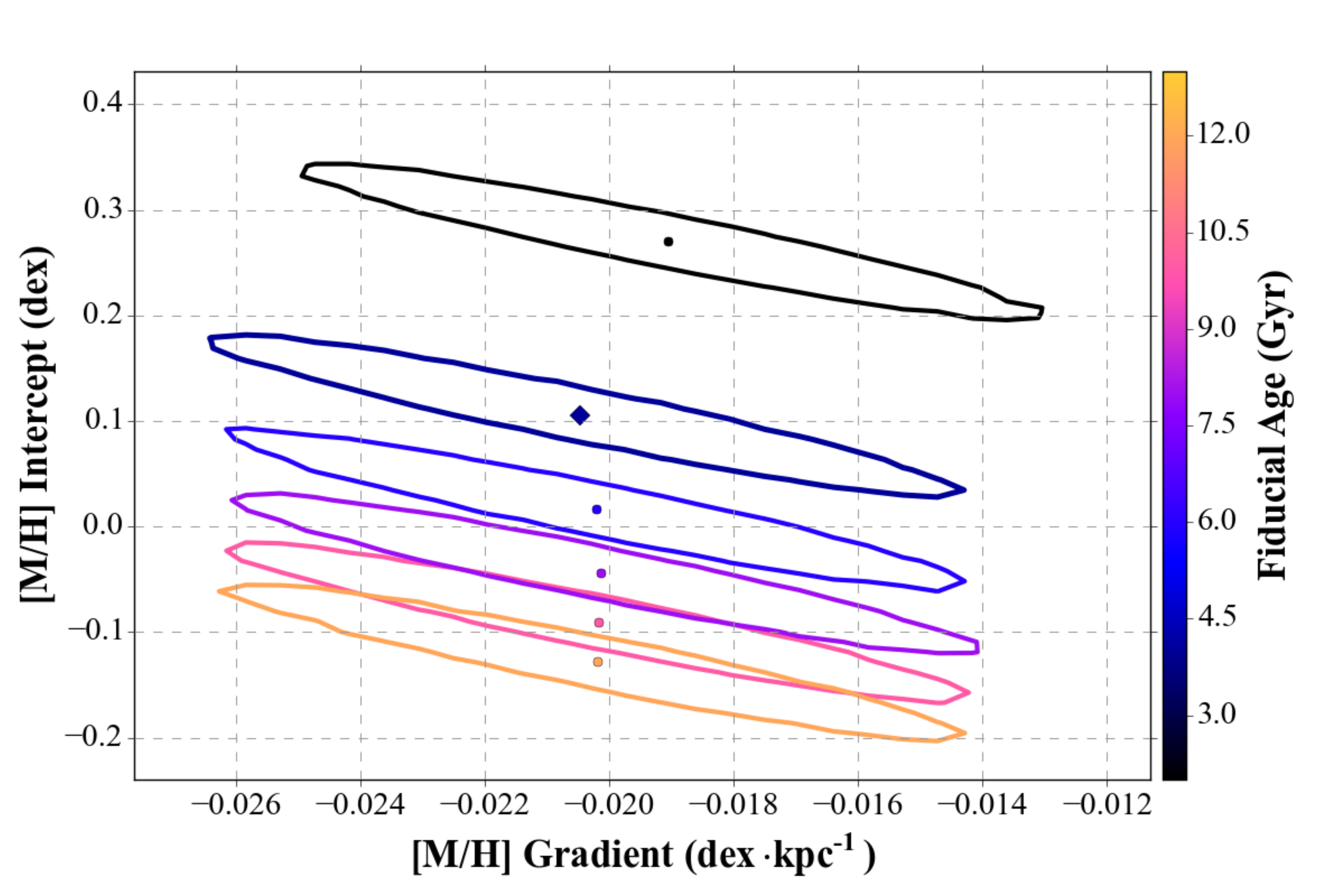}
    \caption{
        This figure shows the best fit slope and intercept for the median metallicity gradient using fiducial ages of 2, 4, 6, 8, 10, and 12~Gyr. The line surrounding each point represents the 1$\sigma$ confidence after performing a Monte Carlo simulation to evaluate the slope and intercept. Our primary result presented in Section~\ref{sec:metallicity-gradient} uses a fiducial age of 4~Gyr shown here as a diamond. The age gradient for all fiducial ages are within 0.25$\sigma$. Changing the fiducial age does have an effect on the absolute metallicity which we discuss in Section~\ref{sec:fiducial-age-effect}. This figure also demonstrates a covariance between the fitted slope and intercept of the metallicity gradient.
    }
    \label{fig:fiducial-age-params} 
\end{figure}

    As discussed in Section~\ref{sec:absolute-metallicities}, the assumed age influences the estimated metallicity of our stars. We evaluate the sensitivity of our metallicity gradient results to changes in the fiducial age by repeating our analysis using isochrone models with ages of 2, 4, 6, 8, 10, and 12~Gyr. For each age, we calculate a distribution of metallicity gradients and intercepts; In Figure~\ref{fig:fiducial-age-params}, we present the best fit and 1$\sigma$ contour of the distribution.

    The metallicity gradient across all fiducial ages is consistent within 0.25$\sigma$. This is because the RGB of isochrones at different ages have similar shapes in the CMD, with the primary difference being that older ages have a lower metallicity at the same RGB color.  This results in a significant change in the normalization of the metallicity gradient and the fitted intercept in metallicity, but little change in the slope.  We find a change in the intercept of roughly $-0.037\pm0.006$~dex/Gyr. 

\subsubsection{The Effect of a Radial Age Gradient}
\label{sec:radial-age-effect}

    A gradient in the age has a significant impact on our metallicity gradient. We evaluate this impact using the following experiment; we fix the age of stars at the center of Andromeda to 4~Gyr. Then, we calculate the age for each star using its spatial position for age gradients of -0.35, -0.25, -0.1, 0.0, and 0.1~Gyr/kpc. We calculate the metallicities similar to in Section~\ref{sec:metallicities} with age in addition to color and magnitude. In this application, the Delaunay triangulation finds the nearest four color, magnitude, and age isochrone points which contain a particular star's parameters and performs the linear interpolation of the metallicity between the points. Using these new metallicities, we recalculate the median metallicity map and derive the distribution of best fit metallicity gradients and intercepts as described in Section~\ref{sec:metallicity-gradient}.

\begin{figure}[ht] 
    \includegraphics[width=\linewidth]{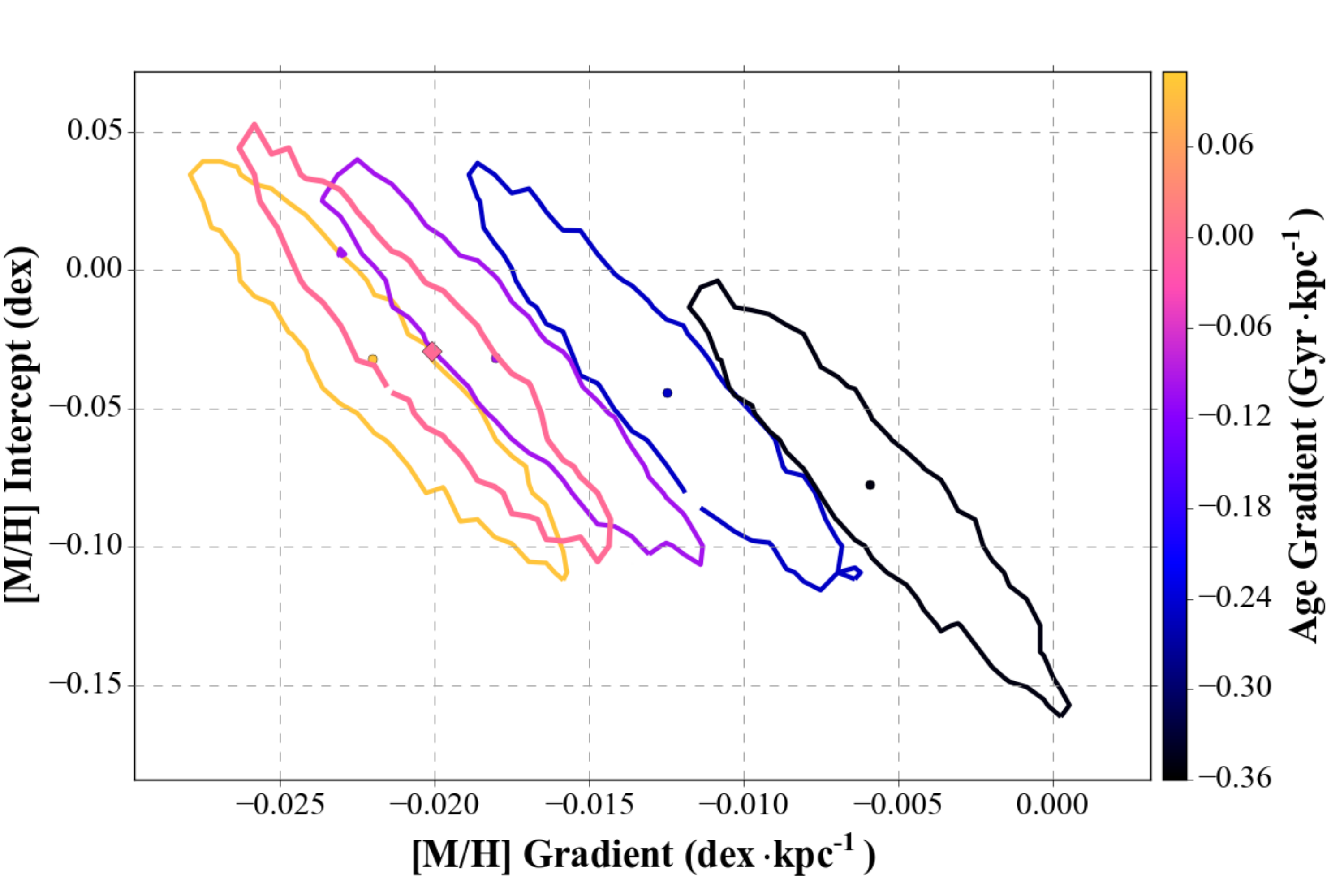}
    \caption{
        This figure shows the effects of an age gradient on the median metallicity. The best fit metallicity gradient and intercept were calculated as described in Section~\ref{sec:metallicity-gradient}, but with a variation of stellar age with radius.  The age gradients shown have slopes of -0.35, -0.25, -0.1, 0.0, and 0.1~Gyr/kpc with a central age of 10~Gyr. The metallicity gradient flattens as the age gradient becomes more negative with it being consistently 0~dex/kpc when the age gradient is -0.35~Gyr/kpc. Based on observations of external galaxies similar to Andromeda, we expect the age gradient to be $\lesssim\arrowvert0.1\arrowvert$~Gyr/kpc (see Section~\ref{sec:comparison-metallicity-gradient}).
    }
    \label{fig:feh-age-vs-radius} 
\end{figure}

    In Figure~\ref{fig:feh-age-vs-radius}, we present the results of this age gradient experiment. As mentioned, as the age gradient becomes negative, the metallicity gradient flattens, while a positive age gradient steepens the gradient. For an age gradient of -0.35~Gyr/kpc the metallicity gradient is consistent with being flat. This figure can be combined with Figure~\ref{fig:fiducial-age-params} to determine the inferred age gradient and intercept for a range of central ages and age gradients.  For instance, if the mean age of stars in our RGB box is 8~Gyr in the center changing by -0.1~Gyr/kpc, then our measured metallicity gradient would be $\sim$-0.018~dex/kpc with a central metallicity of $\sim$-0.04~dex.

    We expect any age gradient in M31 to be small.  This is based on results from the PHAT data (Williams et al. {\em in prep}). Also, \cite{Sanchez-Blazquez2014} find flat mass weighted age-gradients ($< 0.1 $dex/$r_{eff}$) for a majority of galaxies in the CALIFA sample, with the small age gradients being both positive and negative. This translates to an age gradient less than 0.1~Gyr/kpc using values for M31 and our fiducial age. However, resolved star formation histories of M33 and NGC 300 (lower mass, later-type disk galaxies than M31) have shown steeper negative age gradients, with younger ages at larger radii \citep[][]{Williams2009,Gogarten2010}.  If such gradients exist in M31, the younger stars at larger radii will cause a flatter metallicity gradient than using a single fiducial age at all radii. However, resolved star formation histories of M33 and NGC 300 (lower mass, later-type disk galaxies than M31) have shown steeper negative age gradients, with younger ages at larger radii \citep[][]{Williams2009,Gogarten2010}.  If such gradients exist in M31, the younger stars at larger radii will cause a flatter metallicity gradient than using a single fiducial age at all radii.  If any new information becomes available, Figure~\ref{fig:feh-age-vs-radius} can be used to update our metallicity gradient result based on the determined age gradient.


\subsubsection{Comparisons to Literature}
\label{sec:comparison-metallicity-gradient}

    In this section, we compare our measured metallicity gradient to other metallicity gradient measurements in M31 and to metallicity gradients measured in other galaxies.  We find some evidence that the metallicity gradient may become flatter at large radii, but does not appear to vary with age.

       The metallicity gradient of the young population of Andromeda has been measured by \citet{Zurita2012} using HII region abundances over the same radial region covered by the PHAT data.  They find a gradient of $-0.023\pm0.002$~dex/kpc. This gradient is consistent with our RGB metallicity gradient, suggesting that the age gradient has been stable in M31 for some time.

    At larger radii, \citet{Kwitter2012} used abundance measurements of planetary nebula to measure a gradient with deprojected radii of 18 to 43~kpc.  These planetary nebulae have typical ages of 1-2~Gyr, thus probing slightly younger stars than the RGB stars probed here.  They find a gradient of $-0.011\pm0.004$~dex/kpc, significantly flatter than our gradient.  This may be evidence that the metallicity gradient flattens at larger radii.  Another, more indirect approach to measuring metallicities is the stellar population modeling of integrated photometry of M31 from the ANDROIDS survey \citep{Sick2014}.  Their preliminary results seem to show a somewhat shallower gradient of $\sim$-0.01~dex/kpc out to $\sim$30~kpc, with a possible steepening at even larger radii.

    These comparisons provide conflicting support for the expectations due to radial migration \citep[e.g.][]{Roskar2008}.  On one hand, the shallower metallicity gradient in old populations at large radii could be explained if an increasing fraction of stars at large radii come from smaller radii.  However, radial migration also predicts that any metallicity gradient should get shallower over time, thus the similarity of the metallicity gradients in the young and old populations in the inner disk suggests that either the metallicity gradient was steeper in the past, or significant migration has not occurred in the inner disk.

    Higher quality stellar population modeling can be done with spectra, but published results include only the central bulge region.  Spectral index measurements of M31's bulge by \cite{Saglia2010} show the metallicity at radii between 100~pc and 2~kpc is flat and approximately solar.  At smaller radii, the metallicity rises sharply, with [M/H]$\sim$0.5 in the center. At the outer most radii, these measurements include a substantial disk component.  Our central metallicity extrapolated inwards to 2~kpc is roughly consistent with this measurement though the absolute metallicity scale in our measurement is sensitive to many systematics (Figure~\ref{fig:fiducial-age-params} and Section~\ref{sec:absolute-metallicities}). 

    The recent CALIFA survey observed $\sim600$ face-on spiral galaxies using wide-field integrated field units (IFU). Andromeda's size, morphology, and central velocity dispersion are typical among this sample. From the CALIFA sample, \cite{Sanchez-Blazquez2014} selected 62 face on galaxies and measured their stellar metallicities and ages. The mean mass-weighted metallicity gradient of these galaxies is $-0.089\pm0.151$~dex/$R_{eff}$ or $-0.010\pm0.017$~dex/kpc when scaled using M31's effective radius ($R_{eff}=8.9$~kpc; \cite{Courteau2011}). Our metallicity gradient is slightly steeper than this mean metallicity gradient, but is well within the scatter observed of their galaxies. 
    

\begin{figure}[ht] 
    \includegraphics[width=\linewidth]{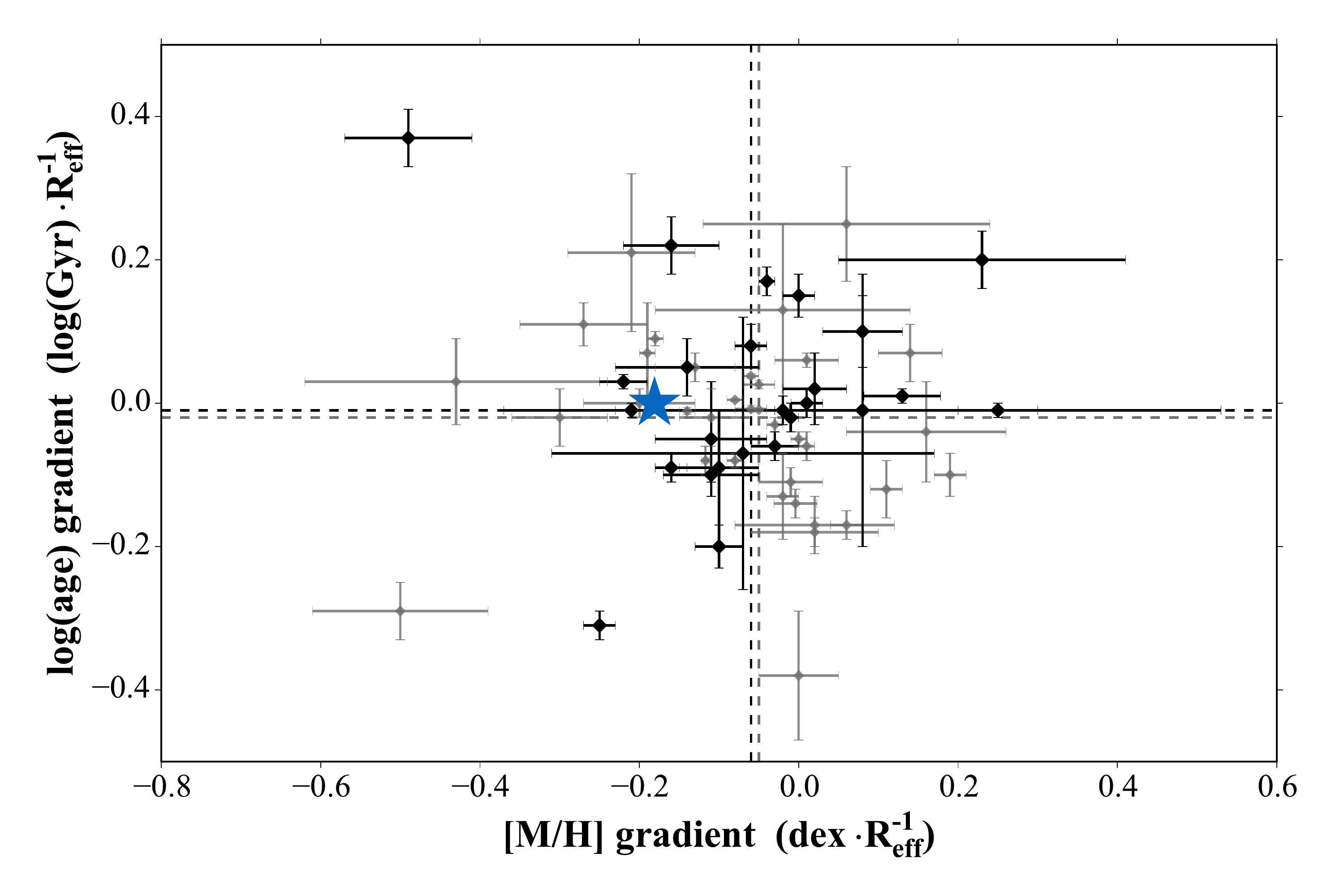}
    \caption{
         Mass-weighted age and metallicity gradients measured for 62 CALIFA galaxies by \citet{Sanchez-Blazquez2014}.  Gray points indicate the full sample, while the 25 galaxies most similar to M31 (Hubble types Sab to Sbc and $-20.5 > M_r > -22.5$) are highlighted in black.  Our M31 metallicity gradient (assuming no age gradient) is shown as the blue star.  Dashed lines give the median of the full sample (gray), and the M31-like sample (black) along both axes.
    }
    \label{fig:califa-comparison} 
\end{figure}

    The metallicity gradient in the Milky Way (MW) disk appears to be significantly steeper than the gradient in M31.  \citet{Cheng2012} measure the metallicity gradient as a function of height (Z) from the disk for SEGUE Survey stars. For stars near the $|Z|=0$ they measure $\sim-0.06$~dex/kpc which drops to $\sim0$~dex/kpc for $|Z| > 1$~kpc (see Figure~8 of \citet{Cheng2012} for metallicity gradient vs Z height). Given that the disk mass is dominated by stars below this height, the MW gradient does appear to be steeper than the M31 gradient. A similarly steep mid-plane gradient is found in other survey data as well including APOGEE data \citep{Hayden2014}, RAVE data \citep{Boeche2014} and the Gaia-ESO survey \citep{Mikolaitis2014}. The shallower gradient in M31 relative to the Milky Way is also discussed by \citet{Kwitter2012}.
    
%

\subsection{Metal Rich Bar Feature}
\label{sec:metal-rich-bar}

    We observe a region of high median metallicity likely associated with the end of the bar. In Figure~\ref{fig:corrected-median-metallicity-map}, this non-axisymmetric region lies just above the major axis between radii of $\sim$3 and 7~kpc. This region contains the highest metallicities found anywhere in our maps. Examination of contours of RGB density in the middle panel of Figure~\ref{fig:corrected-median-metallicity-map} shows a similar axial asymmetry. This asymmetry coincides with M31's bar as measured by \cite{Athanassoula2006} at 8\deg above the major axis. \citet{Davidge2012} also noted an overdensity in AGB star counts near the end of the bar. In addition to the presence of a stellar overdensity, a peculiarly high velocity dispersion ($\sim$150~km/s) is observed at the outer edge of this region \citep{Dorman2015}. This high dispersion could be explained if this represents the end of a misaligned bar.

    This high metallicity region is not caused by dust but represents a real change in the stellar population. As previously noted in Section~\ref{sec:dust-impact}, dust causes the median metallicity to appear more metal rich for moderate extinctions of $A_V\approx0.5$~mag. We observe the high metallicity region prior to the dust correction (Figure~\ref{fig:uncorrected-med-metal}) and it becomes even clearer after the correction. Even more noteworthy, the high metallicity is seen even in the very low extinction regions (dust contours shown in Figure~\ref{fig:corrected-median-metallicity-map}). 
    
    This higher median metallicity region may be expected if either: (1) gas collects at the bar end resulting in higher enrichment there \citep{DiMatteo2013}; or, (2) the bar scatters stars from the inner part of the galaxy which have higher metallicity.  An additional possibility is there could be a shift in the RGB age in the bar region; however, to get a high metallicity region, the average age would need to drop in the bar relative to its surroundings.  Regardless of the mechanism for metal enrichment, this metallicity adds further evidence for a very long bar in M31.

\section{Conclusion}
\label{sec:conclusion}

    We present an analysis of the spatial variations in the metallicity of RGB stars in Andromeda's disk using the PHAT survey. This study is the first large study of RGB metallicities within the Andromeda disk and covers a radial region between 4 and 20~kpc.  Metallicities are estimated for 7.0\e{6} RGB stars using their optical photometry and isochrone models. We create a map of median metallicity and correct this map for the effects of dust extinction and photometric bias and completeness. Our final results are based on the median metallicity map after correcting for these systematics (Figure~\ref{fig:corrected-median-metallicity-map}). From the corrected median metallicity map we conclude: \\
            
\begin{enumerate}

\item Over the inner disk PHAT survey ($\sim$4-20~kpc), we measure a clear gradient in the median metallicity. The measured gradient is [M/H]=$-0.020\pm0.004$~dex/kpc assuming a constant fiducial age of 4~Gyr (Section~\ref{sec:metallicity-gradient}). This gradient has a less than 1~sigma sensitivity to systematic effects of dust extinction, photometric bias and completeness, and fiducial age changes.  The derived metallicity gradient is sensitive to any age gradient that may be present in Andromeda, with negative age gradients resulting in shallower metallicity gradients.  A steep age gradient of -0.35~Gyr/kpc would result in no metallicity gradient; however, spectroscopic observations of other similar galaxies typically show much shallower age gradients \citep{Sanchez-Blazquez2014}.  

\item We measure an enhanced metallicity region offset from the major axis. We show this region is coincident with an overdensity of RGB stars and a region of enhanced dispersion \citep{Dorman2015}.  In Section~\ref{sec:metal-rich-bar}, we discuss how this likely is due to Andromeda's bar \citep{Athanassoula2006}.

\end{enumerate}

\begin{acknowledgments}
\small
We thank the anonymous referee for suggestions that improved the paper.  This work was supported by the Space Telescope Science Institute through GO-12055.  This research made use of codes in the SciPy stack\footnote{{\tt http://www.scipy.org/}}, as well as Astropy, a community-developed core Python package for Astronomy (Astropy Collaboration, 2013)\footnote{{\tt http://www.astropy.org}}. 
\normalsize
\end{acknowledgments}


\bibliographystyle{include/apj} 

\end{document}